\documentclass[aps,prd,twocolumn,superscriptaddress,floatfix]{revtex4-1}
\usepackage{color}
\usepackage{amssymb}
\usepackage[normalem]{ulem}
\usepackage{graphicx,color}
\usepackage{amsmath}
\usepackage{subfigure}
\usepackage[normalem]{ulem}
\usepackage{booktabs}
\usepackage{xcolor}
\usepackage{epsfig,graphics,color,graphicx,amsmath}

\colorlet{darkgreen}{green!50!black}
\colorlet{brightyellow}{yellow!75!red}
\colorlet{orange}{red!50!yellow}
\colorlet{darkblue}{blue!60!black}
\colorlet{darkred}{red!80!black}

\usepackage{soul}

\usepackage{mathtools}
\usepackage{mathrsfs}
\usepackage{bm}
\usepackage{relsize}
\usepackage{indentfirst}

\usepackage{xcolor}

\usepackage{amssymb,amsmath,multirow,epsfig,graphicx,color}

\usepackage{epsfig}
\usepackage{graphicx,amsmath}
\def\be{\begin{eqnarray} &&}

\def\ee{\end{eqnarray}}
\def\psla{\slash \! \! \!}

\def\psla{  \slash \!  \!\!}
\newcommand\ba{\begin{eqnarray}}
\newcommand\ea{\end{eqnarray}}

\newcommand{\bas}{\begin{eqnarray*}}
\newcommand{\eas}{\end{eqnarray*}}

\newcommand{\bno}{\begin{eqnarray*}}
\newcommand{\eno}{\end{eqnarray*}}

\def\psla{ \rlap \slash \! }  

\def\sl
 \usepackage{hyperref}
\usepackage{url}
 \usepackage{hyperref}
\hypersetup{
	colorlinks=true,  
	linkcolor=blue,   
	filecolor=magenta,      
	urlcolor=blue,    
	citecolor=blue,   
} 
\begin{document}
\vspace{-12ex}    
\begin{flushright} 	
{\normalsize \bf \hspace{50ex} LFTC-20-7/59}
\end{flushright}
\vspace{11ex}
 \title{ Exploring the flavor content of light and heavy-light pseudoscalars }
\author{R.~M.~Moita}
\affiliation{Instituto Tecnol\'ogico de Aeron\'autica,  DCTA, 
12228-900 S\~ao Jos\'e dos Campos,~Brazil}
\author{J.~P.~B.~C.~de Melo}
\affiliation{Laborat\'orio de F\'\i sica Te\'orica e 
Computacional - LFTC, 
\\
Universidade Cruzeiro do Sul and 
Universidade Cidade de S\~ao Paulo (UNICID) 
\\  01506-000 S\~ao Paulo, Brazil}
%
\author{ K.~Tsushima} 
\affiliation{Laborat\'orio de F\'\i sica Te\'orica e 
Computacional - LFTC, 
\\
Universidade Cruzeiro do Sul and 
Universidade Cidade de S\~ao Paulo (UNICID) 
\\  01506-000 S\~ao Paulo, Brazil}
\author{T.~Frederico}
\affiliation{Instituto Tecnol\'ogico de Aeron\'autica,  DCTA, 
12228-900 S\~ao Jos\'e dos Campos,~Brazil}

\date{\today}

\begin{abstract}
The  electroweak properties of light and charmed $D$ and $D_s$ pseudoscalar mesons 
are investigated within a unified covariant constituent quark model. 
The quark-antiquark-meson vertices
are assumed to have  a symmetric form by the exchange of quark momenta, 
which is successful in describing 
the light pseudoscalar  meson properties.
The flavor decomposition of the elastic electromagnetic form factors,  
electromagnetic charge radii, and weak decay constants are calculated. 
Based on the results a discussion on the SU(3) and SU(4) symmetry 
breaking is made and a comparison with the pion and kaon
properties to highlight the Higgs contribution to the structure of these mesons.
\end{abstract}
\maketitle

\section{ Introduction} 
\label{intro}

The interplay between dynamical and explicit chiral symmetry breaking in
quantum chromodynamics (QCD), drives the
properties of the heavy-light pseudoscalar mesons, like  $D$ and $D_s$, where dressing 
of the light quarks comes together with the
mass of the heavy partner from the coupling to the Higgs boson. The  consequence of 
the dynamical chiral symmetry breaking 
is the dressing of the light quarks $(u,\,d,\,s)$ and the 
Goldstone boson nature of 
the pion and kaon (see e.g.~\cite{horn2016pion,Aguilar:2019teb}), 
while in heavy sector the charm
quark basically acquires its  mass from the Higgs coupling, breaking badly the SU(4) 
flavor symmetry, separating the Goldstone bosons 
formed by $u\bar d$ and $u\bar s$ from the $c\bar d$ and $c\bar s$ pseudoscalars 
(see e.g.~\cite{Brambilla_2014,Chen_2017}). 

The evolution of the structural properties of the pseudoscalar mesons within the
SU(4) multiplet allows to study the competition between the two
mass generation mechanisms, as the constituent quark masses change from a couple 
of hundreds of  MeV, of the order of $\Lambda_{QCD}$, 
to the GeV scale. Each meson encodes the full
complexity of QCD in Minkowski space, namely its wave function, for example in the 
light-front (LF),  is spread out over an infinite 
set of Fock-components~\cite{bakker2014light}, 
while by itself the dressed quark  degree of freedom encodes such rich structure and it 
is considered a building block, since the primordial era of studies of the 
strong interaction. Nowadays, QCD studies of mesons are far beyond such naive representations 
with several groups performing lattice (LQCD) 
calculations over the world. Also, the dressing of light quarks, gluons and ghosts have been 
computed within LQCD (see e.g. \cite{PhysRevD.99.094506}) strengthening 
the concept of effective quark and gluon degrees of freedom as the
building blocks in 
phenomenological descriptions of hadrons. 
On the other hand the heavy quarks are barely dressed by  gluons, and the Higgs coupling 
being the dominant effect to acquire their masses.

The well separated mass scales of  the  light and the charm quarks  should manifest in the 
heavy-light meson internal structure, as already recognized long ago
(see e.g. \cite{Khlopov1978,Neubert_1994}). The combined study of  mesons, where the largest
component in their wave functions are the non-exotic ones, 
namely a $q\bar q$, formed by a dressed light quark and antiquark or a  heavy-light 
$q\bar q$ pair, should allow to follow the transition in the internal 
structure when a light quark is substituted by a heavy one. In the extreme situation
where the heavy mass tends to infinity, the heavy quark in the $q\bar q$ valence component
is placed at  the center of mass of the meson, while the light quark 
explores the confining 
QCD interaction. The pseudoscalar mesons radically 
change from  the Goldstone boson  nature of the pion and kaon, associated with dynamical 
symmetry breaking, to for example $D$ and $D_s$, where the
chiral symmetry is explicitly broken. Such physical transition should be manifested in the 
structure of these mesons, and in particular in their charge distribution.
In the heavy quark limit part of the charge should be distributed  at the short-range, 
while  an another part at larger distances, while for the pion and kaon, both the quark and antiquark 
should bring somewhat similar charge distributions, apart the individual charge carried by
each constituent.  This sharp modification in the structure of the   light-light to the
heavy-light pseudoscalars  should be reflected in the elastic electromagnetic (EM) form 
factors, and in particular in their flavor decomposition. 

Experimental information on the EM form factors of the pion and kaon are available in 
Refs.~\cite{baldini2000determination,volmer2001measurement,
horn2006determination,tadevosyan2007determination,huber2008charged}
and \cite{Dally:1980dj,Amendolia:1986ui}, 
respectively. Furthermore the charge radii of the pion and kaon are
quite well determined to be, respectively, $r_\pi=\, 0.672\pm0.08$~fm 
and $r_K=\,0.560\pm0.03$~fm (see \cite{Zyla:2020zbs} and references therein).
However, it is still missing experimental information on the
elastic EM form factors of the charged  $D$ and $D_s$ mesons, which would 
be essential to address  the structural modifications moving from 
Goldstone bosons to the heavy-light pseudoscalars. On the other side, 
ab-initio calculations of the $D$ and $D_s$ charge radii at the physical pion mass 
point are not 
yet available, although some results were obtained within a 2+1 flavor
LQCD~\cite{Can2012tx} for pion masses from  300 up to 700~MeV
and with twisted boundary conditions~\cite{2017EPJA} for pion masses
of 300 and 315 MeV. The extracted charge radius  were found around 0.4 fm for the 
$D^+$ and somewhat smaller for the $D^+_s$, indicating the decrease in the size 
of these mesons with respect to the charged pion and kaon, as follows from the 
Higgs coupling to the heavy quarks in opposition to the light ones acquiring 
dynamically their masses. The performed flavor decomposition of the charge radius 
clearly supports the physical picture outlined before.  Additionally, the EM 
form factors and the corresponding flavor decomposition for the $D^+$ up to 
1.5~GeV$^2$~\cite{Can2012tx} and for $D^+$ and $D^+_s$ below 1.2~GeV$^2$ were
computed within LQCD~\cite{2017EPJA}.  It is of note that
the scarce information 
of the EM structure of the pseudoscalar  mesons is  contrasted  by the knowledge of
the weak decay constants from the experiments and LQCD calculations 
(see e.g.~\cite{Zyla:2020zbs}), which is an important piece 
of information of the meson valence wave function at short distances, and necessary
to be taken into account by phenomenological models. 

The above discussion featuring the evolution of the pseudoscalar meson 
structure from
light to heavy-light mesons as represented by their charge distributions,  
motivates our study of $\pi^+,\,K^+,\,D^+$ and $D^+_s$ within a common and 
covariant framework with a minimum number of scale parameters, besides 
the constituent quark masses, all embody in a Bethe-Salpeter (BS) amplitude model.  
It corresponds to the matrix element of an interpolating operator
between the vacuum and the meson state which is built with a minimum number of field
operators characterized by the meson quantum numbers~\cite{itzykson2012quantum}. 
The BS model has a constituent quark and antiquark  and  a pseudo scalar vertex 
with one scale parameter, 
 in a generalization of the model  proposed in~\cite{deMelo:2002yq}, applied with 
 success to compute the pion electroweak properties, and later 
 on used to study the kaon and $D^+$ electromagnetic 
 form factors~\cite{ElBennich:2008qa,Yabusaki_2015}. 
 Furthermore,
the projection of the $q\bar q$ BS amplitude to the LF gives
the valence component of the wave function 
(see e.g. \cite{Frederico:2010zh,MezragFBS}), which allows one to
explore the valence quark momentum distributions 
(see e.g. \cite{2016EPJC...76..253F}
and \cite{de_Paula_2021}).

In the present work, the  BS amplitude model~\cite{deMelo:2002yq} is applied to 
compute the EM form factors  of  $\pi^+,\,K^+,\,D^+$ and $D^+_s$, as well as 
their flavor decomposition, 
via the Mandelstam formula~\cite{Mandelstam:1955sd},
represented by the triangle Feynman diagram. The model has constituent quarks $u$, $d$, $s$ 
and $c$, 
with fixed masses and one individual scale parameter fitted to the well known value of 
each meson
decay constant. The model is covariant and conserves the EM current, as the constituent 
quarks are point like, with a bare current, which trivially satisfy the Ward-Takahashi 
identity~\cite{itzykson2012quantum}. In addition, the decay constant is computed from 
the antialigned quark spin component of the LF valence wave function, which is
derived from the BS amplitude model.

In sect.~\ref{sec:model}, an analytical form for the Bethe-Salpeter amplitude in terms of 
constituent quarks for the pseudoscalars,  $\pi^+,K^+,D^+$ and $D^+_s$  
is proposed within a unified covariant model, and 
from that the weak decay constant is derived, and its association with the 
antialigned quark spin component of the valence LF
wave function is presented. In sect.~\ref{sec:em}, the electromagnetic current 
for the elastic process is constructed, the flavor decomposition of the elastic 
electromagnetic form factors 
is derived, and the method for treating numerically the loop integrations 
with LF technique is discussed. 
The results for the static electroweak observables are provided in sect.~\ref{sec:SEW}, 
and discussed in  comparison with LQCD calculations and  other models. The electromagnetic
form factors from our model are discussed in sect.~\ref{sec:EMFF} and compared 
with the vector meson dominance model and with experimental 
data for the pion and kaon, while for $D^+$ and 
$D^+_s$ with LQCD results. 
The work is closed in sect.~\ref{summary} with a summary of the
main results.

\section{The covariant framework} 
\label{sec:model}

\subsection{Quark-meson spin coupling: effective Lagrangian}
We adopt here a simple scheme to build the spin coupling of the quark-antiquark
pair to build the meson starting from an effective Lagrangian. 
Note that, later on 
a  meson vertex will be introduced carrying a mass scale dictated
by the weak decay constant.

We start by coupling the quark to the pseudoscalar meson field within the SU(4) 
flavor symmetry scheme, which is expressed by the following effective Lagrangian:
\begin{equation}\label{eq:Leff}
 {\cal L_I} = -\imath g\, \bar \Psi M_{\rm SU(4)} \gamma^5\Psi\equiv -\imath \, 
 \frac{g}{\sqrt{2}} \sum_{i=1}^{15}
 (\bar \Psi \lambda_i \gamma^5\Psi) \varphi^i\, ,
 \end{equation}
where $g$ is a coupling constant,  $\lambda_i \,(i=1, ... ,15)$ are the SU(4) 
Gell-Mann matrices~\cite{Close:1979bt},
$\varphi^i$ is the Cartesian components of the pseudoscalar meson fields,
the quark field is
$\Psi^T=(u,d,s,c)$ ($T$: transposition)
decomposed in its quark-flavor components and
\begin{widetext}
 \begin{equation}   
M_{SU(4)}  =  \ 
\frac{1}{\sqrt{2}}
\left( \begin{array}{llll}
\frac{\pi^0}{\sqrt{2}} + \frac{\eta}{\sqrt{6}} + \frac{\eta_c}{\sqrt{12}}
	    & \ \ \pi^+    &  \ \ K^+     & \ \ \bar{D}^0 \\
\pi^-   &  \ \ -\frac{\pi^0}{\sqrt{2}} + \frac{\eta}{\sqrt{6}}
 + \frac{\eta_c}{\sqrt{12}} & \ \ K^0   & \ \ D^- \\
K^-    & \ \ \bar{K}^0   & 
\ \ -\sqrt{\frac{2}{3}} \eta + \frac{\eta_c}{\sqrt{12}}  & \ \ D^-_s \\ 
 D^{0} & \ \ D^{+}    & \ \ D^{+}_s & \ \ -\frac{3 \eta_c}{\sqrt{12}} 
\end{array} 
\right) ~,
\end{equation}
\end{widetext}
is the SU(4) pseudoscalar meson  field matrix 
\cite{Lin:1999ve,Bracco:2011pg,Georgi:1985kw}.

In particular, the positively charged pseudo-scalar mesons which we focus in this 
study are selected from the SU(4) meson matrix through the traces 
\begin{small}
\begin{eqnarray}
& &\pi^+=Tr\left[M_{SU(4)}\,   \lambda_{\pi^+}\right],
\,K^+=Tr\left[M_{SU(4)}\,   \lambda_{K^+}\right]\, ,
\nonumber \\
& & D^+=Tr\left[M_{SU(4)}\,   \lambda_{D^+}\right] , 
\,
D^{+}_s=  Tr\left[M_{SU(4)}\,   \lambda_{D^+_s}\right],
\end{eqnarray}
\end{small}
where the flavor matrices are given by:
\begin{eqnarray}
& &\lambda_{\pi^+}= \frac{1}{\sqrt{2}}(\lambda_1 + \imath \lambda_2),
\, \lambda_{K^+}= \frac{1}{\sqrt{2}} ( \lambda_4 + \imath \lambda_5 ),
\nonumber \\
& & \lambda_{D^+}=  \frac{1}{\sqrt{2}}(\lambda_{11} - \imath \lambda_{12}),
\,
\lambda_{D^{+}_s}=  \frac{1}{\sqrt{2}}(\lambda_{13} - \imath \lambda_{14}),
\end{eqnarray}
and the corresponding physical mesons are indicated by the 
subindices.

\subsection{Bethe-Salpeter amplitude model}
The effective Lagrangian from Eq.~\eqref{eq:Leff}, is associated to a 
meson vertex without structure and point-like, introduced only to guide us in a 
practical form to build both the spin and favor composition of each meson. In what follows,
we will allow the meson vertex to have an extension, represented by a scalar 
function to keep the  covariance of the model. In this way,  
the Bethe-Salpeter amplitude model 
for the pseudoscalar mesons considered in this study is given by: 
\begin{small}
\begin{equation}
\Psi_M(k,p) = S_q\left(k\right)
\gamma^5 g\Lambda_M(k,p) \lambda_M S_{q}\left(k-p\right)
~,
\label{bsalpeter1}
\end{equation}
\end{small}
where the constituent quark propagator is
\begin{equation}
S_q(k)=i[\,\psla{k} - \widehat m_q+ \imath\epsilon\,]^{-1} \ , 
\end{equation}
and the quark constituent mass matrix is diagonal, 
$${\rm diag} 
[\widehat{m}_q]=[m_u,~m_d,~m_s,~m_c].$$
The vertex function  for the pseudoscalar mesons,
$M=(\pi,K^+, D^+,D^+_s)$, adopted
in the present work is
\begin{equation}\label{vertexmom}
g \Lambda_M(k,p)= \frac{C_M}{k^2-\mu^2_M + \imath \epsilon } + 
[k\to p-k]
 \, ,
\end{equation}
which generalizes the model proposed in Ref.~\cite{deMelo:2002yq} for the pion and the
kaon~\cite{Yabusaki_2015}  to the heavy-light case.
The model assumes that the infrared (IR) dynamics of QCD is translated to the  
mass scale, $\mu_M$, 
for each pseudoscalar meson in the SU(4) flavor multiplet. The ultraviolet (UV) 
physics is reflected in the
analytic form of the vertex function.
The constant $C_M$ also depends on the meson and it is determined by 
the covariant normalization of the BS amplitude:
\begin{eqnarray}
&&
2ip^\mu=N_c \,\text{Tr}\int \frac{d^4k }{(2 \pi)^4} g^2 \Lambda^2_M(k,p) 
\nonumber	\\ && \times  \Bigg[ 
		\gamma^5 \lambda_M S_q(k- p) 
		  \gamma^\mu S_q( k-p)
	\gamma^5\lambda^\dagger_M S_q(k)
	\hspace{-0.1cm}  \nonumber
	\\ && +
		\gamma^5 \lambda^\dagger_M S_q(k+p) 
		 \gamma^\mu S_q( k+p)
	\gamma^5\lambda_M S_q(k)
	\Bigg] \, ,
\end{eqnarray}

where it was made the simplified assumption that the kernel which would have given
origin to this particular vertex function had no 
dependence on the total momentum, as it is the case of the ladder approximation 
of the BS equation (see e.g. \cite{Maris:2000sk,de_Paula_2021}).

The breaking of the SU(4) symmetry is reflected in the variation of the mass scale  
$\mu_M$ and the constituent quark masses 
as a consequence of both mass generation by the Higgs mechanism and the
dynamical chiral symmetry 
breaking. In particular, $\mu_M$ is obtained 
by fitting $f_M$, the weak decay constant of the meson $M$, for a given set
of constituent quark masses.

We observe that the masses of the constituent quarks are associated
to an energy scale characteristic of each meson.
Such energy scale sets the initial condition for the evolution to obtain 
the parton distribution function at the different energy scales. For practical applications,
it is about 0.5~GeV for the pion (see e.g.~\cite{2016EPJC...76..253F}, 
but could change with the meson.  

Another comment is appropriate, in order to keep the simplicity of the present phenomenological 
covariant model, we have adopted the same form of the vertex function for all mesons,
which at large momentum 
behaves as $1/k^2$. Such asymptotic form should naively correspond to the 
situation where the quark and antiquark, exchange a very large momentum, 
flowing through the one-gluon exchange interaction, that due to the
asymptotic freedom dominates the short-distance dynamics of the 
system~(see e.g. Ref.~\cite{LepPRD1980}).
Of course, we could have  other types of vertices, at the expense of introducing 
more parameters, but we chose to keep the minimal number of scale parameters in this work, namely one per meson.

\subsection{Weak decay constant}

The pion weak decay constant is a measure of the strong interaction dynamical scale, 
and as such 
a fundamental requirement that a model satisfies. The weak decay 
constant comes as a balance of both  short-range and long-range QCD physics
to the meson valence wave function, and therefore a necessary constrain in
phenomenological models. 
In the present work, the chosen model satisfies such physical requirement not only
for the pion but 
also for all the pseudoscalar mesons. The pseudoscalar meson decay constants encode
relevant physical information on the structure of the pseudoscalar mesons, allowing together
with the Cabibbo-Kobayashi-Maskawa~(CKM) matrix elements~\cite{Zyla:2020zbs,Ablikim:2018jun} 
via the leptonic weak decay,
$M\rightarrow l \nu_l$ ($l$ represents the charged leptons, $l=e,\mu,\tau$) to obtain the
weak decay width. In the lowest order it is given by~\cite{Zyla:2020zbs,Ablikim:2018jun} :
\begin{equation}
\Gamma(M \to l\nu_l ) \hspace{-.1cm}= \hspace{-.1cm}\frac{G^2_F}{8 \pi }
f^2_{M} m^2_l m_{M}\hspace{-.1cm}
\left( \hspace{-.1cm}1- \frac{m^2_l}{m^2_M}\hspace{-.1cm}\right)^2  |V_{q_1 q_2}|^2 ,
\end{equation}  
where,~$ G_F$~is the Fermi coupling constant, $m_l$ is 
the lepton mass, $m_M$ is the pseudoscalar meson mass, 
and $V_{q_1 q_2}$ is the corresponding CKM
matrix element.

The pseudoscalar meson decay constant, $f_M$, 
is defined through the matrix element of the axial-vector current
operator~\cite{Salcedo:2003yb,itzykson2012quantum}, 
\begin{equation}
\langle 0| A^j_\mu | M^k \rangle  =  \imath p_\mu   f_{M^k} \delta^{jk},
\end{equation} 
where  $ A^j_\mu = \bar{q}(0) \gamma_\mu \gamma^5 \frac{\lambda_j}{2} q(0)$ is
the axial-vector current.
The indices $j$ and $k$ identify the 
isospin (flavor) components of the current operator  
and pseudoscalar meson.   

\begin{figure}[t]
\begin{center}
\epsfig{figure=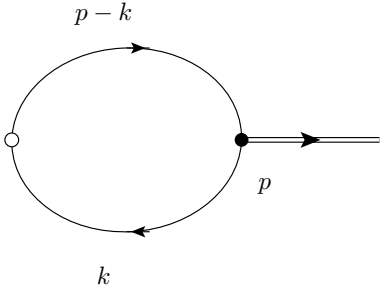 ,width=5.0cm}
\caption{Diagrammatic representation of the pseudoscalar meson weak decay amplitude.}
\label{fig2}
\end{center}
\vspace{3ex}
\end{figure}

According to the diagram shown in Fig.~\ref{fig2}, we obtain 
the following expression for the decay constant, with
the vertex function from Eq.~\eqref{vertexmom}:
\begin{multline}
  \imath \,p^\mu f_{M}  = 
  N_c  \int \frac{d^4k}{(2 \pi)^4} 
 \frac{1}{2} 
  \text{Tr}\left[ 
\gamma^\mu \gamma^5  \lambda^{\dagger}_M
\Psi (k,p)
\right] 
\,,
\label{fM}
 \end{multline}
where the pseudoscalar meson is simply labeled by $M$ and the trace is taken over the spinor
and  flavor spaces.
$N_c=3$ is the number of quark  colors. 

The decay constant $f_M$ in~\eqref{fM} is evaluated in the rest frame of the pseudoscalar 
meson, $p^\mu=(m_M,\vec{0})$, considering the plus component of 
the  axial-vector current corresponding to 
$\gamma^+\gamma^5=(\gamma^0+\gamma^3)\gamma^5$, 
and the loop integration is performed with  LF momentum.
After integration over the LF energy $k^-$, we obtain:
 \begin{multline}\label{fdecay}
 f_{M}   = 
 \,\frac{N_c }{4 \pi^3} 
 \int d^2k_\perp
 \int_0^1 dx\,
 \psi_M(x,\vec k_\perp;m_M,\vec{0}_\perp),
 \end{multline}
 where $\psi_M$  is the momentum part of the antialigned quark spin component of the 
 pseudoscalar meson 
 valence wave function, given by:
\begin{multline}\label{PSILF}
 \psi_M(x,\vec{k}_\perp;p^+,\vec{p}_\perp) = \frac{p^+}{m_M}  
 \frac{g\, C_M}{m^2_M - {\cal M}^2(m_q,m_{\bar{q}})}
\\
\hspace{-2cm}\times  \Bigg[  \frac{1}{(1-x) (m^2_M - {\cal M}^2(m_q,\mu_M) )} 
  \\ +  \frac{1}{ x (m^2_M - {\cal M}^2(\mu_M,m_{\bar{q}}) }
  \Bigg]
   \\  +\left[ m_q\leftrightarrow m_{\bar q}\right], 
  \end{multline}
where, $x=\frac{k^+}{p^+}$,~$0 < x < 1$, and   
\begin{multline}
 {\cal M}^2(m_1,m_2)= \frac{|\vec{k}_\perp|^2 + m^2_1}{x} \\ + 
 \frac{|\vec{p}_\perp-\vec{k}_\perp|^2 + m^2_2}{1-x} 
 - |\vec{p}_\perp|^2 
 \, .
\end{multline}

Note that, the valence wave function is obtained 
from the Bethe-Salpeter amplitude~(\ref{bsalpeter1}), 
by integration over the light-front energy,~$k^-$, after the instantaneous terms  of the
quark propagators are dropped out
(see Ref~\cite{Frederico:2010zh} for more details).  
The plus component of the axial-vector current in \eqref{fM}
due to the property $(\gamma^+)^2=0$  kills the instantaneous terms of the quark propagator and 
the choice of $\gamma^+\gamma^5$ to obtain the  decay constant
selects the valence wave function with antialigned quark 
spins (see Refs.~\cite{MezragFBS,de_Paula_2021}).

\section{Electromagnetic current}
\label{sec:em}
The quark electromagnetic current operator for the photo-absorption process 
in a point-like constituent quark is defined by,
\begin{equation}
j^\mu = \frac{2}{3} \bar{u} \gamma^\mu u
+ \frac{2}{3} \bar{c} \gamma^\mu c 
 - \frac{1}{3}
\bar{d} \gamma^\mu d 
-\frac{1}{3}  \bar{s} \gamma^\mu s
\label{jmu}
\end{equation}
where $u$, $d$, $s$ and $c$ are the quark fields. In the flavor space  
$${\rm diag}~[\widehat{Q}]=[e_u,~e_d,~e_s,~e_c]=[2/3,~-1/3,~-1/3,~~2/3]\, ,$$
is the charge operator  defined by a diagonal matrix.

\begin{figure}[t]
\begin{center}
\epsfig{figure=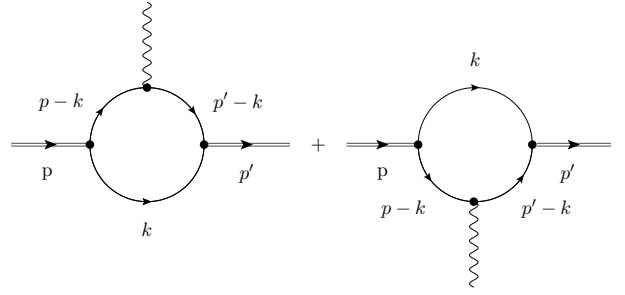,width=8.0cm} 
\caption{Feynman diagrams representing 
the electromagnetic interactions 
with pseudoscalar mesons for calculating the elastic
electromagnetic form factors
in the present work, where expressions for each meson 
are given in Eq.(\ref{ffseparate}).
}
\label{fig1}
\end{center}
\end{figure}

The matrix element of the electromagnetic current for each meson is obtained
from the Mandelstam formula represented by the
Feynman diagrams depicted in Fig.~\ref{fig1}
(c.f. Hutauruk et al.~\cite{Hutauruk:2016sug}):

\begin{eqnarray}
&& \langle p';M |j_1^\mu |p;M\rangle	  = -i N_c
\int \frac{d^4k }{(2 \pi)^4} g \Lambda_M(k,p') g \Lambda_M(k,p)
\nonumber	\\ && \times \text{Tr} \left[ 
		\gamma^5 \lambda_M S_q(k- p) 
		 \widehat{Q} \gamma^\mu S_q( k-p^{\prime})
	\gamma^5\lambda^\dagger_M S_q(k) 
	\right]\hspace{-0.1cm} ,
\nonumber	\\
&& \langle p';M |j_2^\mu |p;M\rangle	  =-i N_c
\int \frac{d^4k }{(2 \pi)^4} g \Lambda_M(k,p') g \Lambda_M(k,p)
\nonumber	\\ && \times \text{Tr} \left[
		\gamma^5 \lambda^\dagger_M S_q(k+p') 
		 \widehat{Q} \gamma^\mu S_q( k+p)
	\gamma^5\lambda_M S_q(k)
	\right]\hspace{-0.1cm} ,
\label{jcurrent}
\end{eqnarray}
 where the trace is performed over
the Dirac and flavor indices, and the total microscopic current is
\begin{small}
\begin{equation}
\langle p^\prime;M | j^\mu| p;M \rangle = 
\langle p^\prime;M | j_1^\mu| p;M \rangle \hspace{-0.1cm} +\hspace{-0.1cm} 
\langle p^\prime;M | j_2^\mu| p;M \rangle \, . 
\end{equation}
\end{small}
Note the above expression  contains the
two diagrams shown in Fig.~\ref{fig1}, and they are  written 
in flavor space and represent the photon being absorbed by each quark of the
meson.

The space-like elastic electromagnetic form factor  is 
extracted  by equating  the covariant expression
\eqref{jcurrent} to the macroscopic formula of the current:
\begin{equation}
\langle p^\prime;M | j^\mu| p;M \rangle = 
(p^{\prime \mu}+p^\mu ) F_M (q^2)~, 
\label{ffactor1}
\end{equation}
where $q=p'-p$ is the momentum transfer and  $F_M (q^2)$ is 
the elastic electromagnetic  
form factor. 

We point out that the normalization constant~$C_M$ appearing in the 
vertex function \eqref{vertexmom} is 
determined by $F_{M}(0)=1$, i.e. the form factor normalization.
In addition, the mass scale, $\mu_M$,  is fitted  to reproduce the experimental weak decay constant.
The constituent quark mass values ($m_q$) are chosen according to
previous studies ~\cite{Suisso:2002jg,deMelo:2002yq,Yabusaki_2015} with similar covariant models.

\subsection{Flavor decomposition}

We can separate the individual contribution from each quark in 
Eq.~\ref{jcurrent} by writing the two traces as the sum of two terms: 
one associated with the photon being absorbed by the quark 
with charge +2/3; and  other one corresponds  to the antiquark with
charge +1/3. Therefore, we have that for the sum of the two traces:
\begin{equation}
\text{Tr}[ \ ]= 2 \left( \frac{2}{3} \Delta^\mu_{a\bar ba} + 
\frac{1}{3} \Delta^\mu_{\bar b a \bar b} \right), 
\label{delta12}
\end{equation}
which can be rewritten simply as:
\begin{equation}
\Delta^\mu_{aba}= 
                \text {Tr} \left[ 
		\gamma^5 S^a_{q}(k- p) 
		  \gamma^\mu S^a_{q}(k-p^{\prime}) S^b_{q}(k) \gamma^5 \right],
\end{equation}
the\, same\, as\, \eqref{jcurrent}\, $j^\mu_1$, 
where the diagonal matrix element of the quark propagator is $S^a_{q}$ for
flavor $a$. The photon probes in each case the quark or antiquark 
labeled $a$ in the first term
and $\bar{b}$ in the second term of Eq.~\eqref{delta12}, 
 corresponding to the quarks ($u$ or $c$)  and to the antiquarks 
 ($\bar d$ or $\bar s$), respectively.

Therefore, the matrix element of the current can be decomposed in 
the quark flavor content according to:
\begin{eqnarray}
 \langle p';M |j^\mu |p;M\rangle	  &=& -2i N_c\hspace{-.1cm}
\int \frac{d^4k }{(2 \pi)^4}	 g\Lambda_M(k,p') g\Lambda_M(k,p)
\nonumber	\\ && \times \left( \frac{2}{3} \Delta^\mu_{a\bar ba} + 
\frac{1}{3} \Delta^\mu_{\bar ba\bar b} \right) ,
\label{jcurrentflavor}
\end{eqnarray}
where the photon interacts with the quark $a$ in the first term and with 
the antiquark $\bar{b}$ in the second one.
From that we can write the flavor decomposition of the form factors
\begin{eqnarray}
F_{\pi^+} (q^2) & = & {\frac{2}{3} F_{u \bar{d} u}(q^2) + 
\frac{1}{3} F_{\bar{d} u \bar{d}}}(q^2)\, ,
\nonumber \\
F_{K^+} (q^2) & = & { \frac{2}{3} F_{u \bar{s} u}(q^2)  + 
\frac{1}{3} F_{\bar{s} u \bar{s}}}(q^2)\, , 
\nonumber \\
F_{D^+} (q^2) & = & { \frac{2}{3} F_{c \bar{d} c}(q^2)  + 
\frac{1}{3} F_{\bar{d} c \bar{d}}(q^2)\,,} 
\nonumber \\
F_{D^+_s} (q^2) & = & { \frac{2}{3} F_{c \bar{s} c }(q^2)  +
\frac{1}{3} F_{\bar{s} c \bar{s}}(q^2)\, .}
 \label{ffseparate}
\end{eqnarray}
In the SU(2) isospin symmetry limit  
with the $u$ and $d$ quark masses being equal,  
Eq.~(\ref{jcurrentflavor}), implies  that
$F_{\pi^+}(q^2)= 
F_{u \bar{d}u}(q^2)= F_{\bar{d}u\bar{d}}(q^2)$. For $K^+$, $D^+$ 
and $D^+_s$, respectively the SU(2), S(3) and SU(4) symmetries are broken.
By the charge conservation, it is required that
\begin{eqnarray}
&&F_{u  \bar{s}u}(0) \ = \ F_{\bar{s}u\bar{s}}(0)  =1\, , \nonumber
\\
&&F_{c \bar{d} c}(0) \ = \ F_{\bar{d} c \bar{d}}(0) =1\, ,  \nonumber
\\ 
&& F_{c \bar{s} c }(0) \ = \ F_{\bar{s} c \bar{s}}(0)  =1\, . \label{ffnorm}
\end{eqnarray}
The partial quark contributions to each  meson form factor becomes different
 when increasing the momentum transfer, despite the same normalization, as will 
be shown by our calculations.

\subsection{LF  technique}

The calculation of the elastic photo-absorption transition amplitude, 
Eq. \eqref{jcurrentflavor}, is performed in the
Breit frame, with the choice of initial and final meson four-momentum 
$p^\mu = (\sqrt{m^2_M+\tfrac14 q_x^2},-\tfrac12 q_x,0,0)$ and
$p^{\prime \mu }=(\sqrt{m^2_M+\tfrac14 q_x^2},\tfrac12 q_x,0,0)$, respectively. Furthermore,
the meson light-front momentum components are chosen as  
$p^+=p^{\prime +}=p^-=p^{\prime -}$,
$p'_\perp=( q_x/2,0)$ and $p_\perp=(-q_x/2,0)$, which 
corresponds to $q^+=0$ fulfilling the Drell-Yan 
condition~\cite{Brodsky:1997de}.
The form factor is obtained from   the plus component 
of the EM current $J^+=J^0+J^3$, implying in the usage of the quark
current associated with
$\gamma^+=\gamma^0+\gamma^3$ in Eq. ~\eqref{jcurrentflavor}, 
when the relevant Dirac trace is performed,  to give:
\begin{equation}
F_M(q^2)  = \frac{1}{2p^+}\langle p' | j^+_{M} | p \rangle~.
 \label{ffactor2}
\end{equation}
We should stress that the choice $\gamma^+$ eliminates the 
instantaneous terms of the fermion propagators attached to the 
quark EM current.
The loop integration is carried out analytically over $k^-$, 
the light-front energy, 
and, after the integrations over $k^+$ and $k_\perp$ are performed. 
Relevant to observe that the choice of the Drell-Yan frame and
plus component of the current is enough to eliminate the end-point
singularities for this pseudoscalar model 
(see \cite{deMelo:1997cb,deMelo:2002yq}).
However, for 
frames with $q^+ \neq 0$, in order to preserve the full covariance of the model,
it is necessary to take into account a nonvalence contribution to the form 
factor~\cite{Bakker:2000pk,deMelo:2002yq}.

As a technical remark,  the Feynman parameterization could
be used alternatively to evaluate the one-loop integrals, keeping the explicit
covariance of the model at all steps of the calculations.
 For our purpose, using the light-cone variables, as we did, or Feynman
 parametrization should not  affect the quantitative results.
\begin{table*}[hbt!]
\begin{center} 
\caption{Pseudoscalar meson static electroweak  observables. The notation for the entries,   
(A,B,C,D,E,F) respectively correspond to the different model parameters 
and pseudoscalar mesons for ($\pi^+,\pi^+,K^+,K^+,D^+,D^+_s$). In particular, 
the results with models B and D are from Ref.~\cite{Yabusaki_2015}. 
 Note that, all the relevant constituent quark mass values for the models A, C, E, 
and F, are from Ref.~\cite{Suisso:2002jg}.
The experimental data come from Refs.~\cite{Zyla:2020zbs,Ablikim:2018jun}. 
The  masses $m_q, m_{\bar{q}}$,  $\mu_M$, 
binding energy $(\epsilon_M)$ and decay constant $(f_M)$ are given  in [MeV]. The
charge radius $(r_M)$ is given in [fm].
}
\bigskip
\begin{tabular}{|c|c|c|c|c|c|c|c|c|c|c|c|}
\hline
\hline
Model/meson& Flavors   & $I(J^P)$ & $m_q$    &  ~$m_{\bar{q}}$ & $m_M$ & 
$\epsilon_M$ & $\mu_M$ & ~$r_M$ &~$f_{M}$ 
& ~$r_{M}^{\rm Expt.}$ &  ~$f^{\rm Expt.}_M$ \\
\hline 
 (A) $\pi^+$  & $~u\bar d$ &$1(0^-)$   & ~384  & ~384& 140 &628  &~225  &~0.665 & ~92.55  & ~$0.672(8)$ &
~$92.28(7) $ \cite{Zyla:2020zbs}   \\
(B)~~~~~    &     &   & ~220  & ~220  &  & 300&~600  &~0.736 & ~92.12  &              &        
\\
\hline
(C)      $K^+$    &  $~u\bar s$ &$\frac12(0^-)$     &  ~384 & ~508 & 494& 398 &~420 
&~0.551 &~110.8  & ~$0.560(3)$ &~$110(1)$ \cite{Zyla:2020zbs}       \\
(D)~~~~~~    &    &      & ~220  & ~440  & & 166 &~600  & 0.754  &~110.8  &              &  
\\
\hline
(E)  $D^+$    &    $~c\bar d$  &$\frac12(0^-)$     & ~1623  & ~384 &  1869 & 138 
&~1607  &~0.505 &~144.5  &    &~$ 144(3) $ \cite{Zyla:2020zbs}         \\
\hline
(F) $D^+_s$   &  $~c\bar s$   &$0(0^-)$      & ~1623  & ~508  & 1968 &163 &~1685 &~0.377 &~182.7
 &         &~$182(3)$  \cite{Zyla:2020zbs}       \\
%
 &  &  & & & & & & & & & ~$179(5)$ \cite{Ablikim:2018jun} \\
\hline
\hline 
\end{tabular}
\label{table1}
\end{center}
\end{table*}
 
 \begin{table*}[htb]
 	\begin{center}
 		\caption{
Decay constants and electromagnetic radii of $\pi^+$ (model A) and $K^+$ (model C)
in the present model, compared with the other works in the literature, 
as well as the experimental data in particle data group (PDG)~\cite{Zyla:2020zbs}. 
The decay constants are in [MeV], and the charge  
radius are in [fm].}
\label{table3}
\begin{tabular}{|c|c|c|c|c|c|} 
\hline
\hline
 Reference                   &   $f_\pi^+$    & $f_K^+$  & ~$r_{\pi^+}$ & ~$r_{K^+}$ &~$f_{K^+}/f_{\pi^+}$ \\ 
\hline 
This work                    & 92.55         & 110.8    & 0.665     & 0.551  &  1.196   
\\
Maris \& Tandy~\cite{Maris:2000sk} 	  & 92.62 & 109.60 & 0.671 & 0.615 & 1.182 
 \\ 
Faessler et al.~\cite{Faessler2003} & 92.62 & 113.83  &   0.65     &       &   1.23  
 \\
Ebert et al.~\cite{ebert2006,ebert2005}  & 109.60  & 165.45      &  0.66  &  0.57 & 1.24  \\
Bashir et al.~\cite{Bashir:2012fs} & 101    &        &            &       & \\
Chen \& Chang~\cite{Chen:2019otg}   & ~93    & ~111    &          &       & 1.192   \\
Hutauruk et al.~\cite{Hutauruk:2016sug} & 93    &  97   & 0.629 & 0.586 &  1.043 \\
Ivanov et al.~\cite{Ivanov:2019nqd}  & 92.14  & 111.0   & &  &  1.20 \\
Silva et al.~\cite{daSilva:2012gf} & 101    & 129    & 0.672      & 0.710 & 1.276 \\
Jia \& Vary~\cite{Jia:2018ary}       & 142.8  & 166.7   &0.68(5)   & 0.54(3) & 1.166 \\
 \hline
PDG~\cite{Zyla:2020zbs}&$92.28(7)$ &$110(1)$ &0.672(8) & 0.560(3) & 
 1.192(14) \\
 			\hline 
 			\hline
 		\end{tabular}
 		\label{table2}	
 	\end{center}
 \end{table*}

 \section{Static Electroweak observables }\label{sec:SEW}

The Bethe-Salpeter amplitude model 
for the $\pi$, $K$, $D$ and $D_s$ has for each meson three parameters: the
constituent quark masses $m_q$ with $q$ from $\{u,\,d,\,s,\, c\}$ and a mass 
scale $\mu_M$ (see Eqs.\eqref{bsalpeter1} and \eqref{vertexmom}). The parameter
$\mu_M$ constrains the model to provide the observed weak decay constant.
We work here with six sets of parameters, namely
(A,B,C,D,E,F), which respectively correspond to the pseudoscalar mesons
($\pi^+,\pi^+,K^+,K^+,D^+,D^+_s$), as well as to the different 
choices of quark masses as given in Table~\ref{table1}.
The choices of constituent  quark masses are: for the light quark  mass values  
of 384 MeV  as estimated in~\cite{Suisso:2002jg}  (386 MeV 
was used in~\cite{Suisso:2002jg}) and 220 MeV~\cite{deMelo:2002yq};
the strange constituent  mass values of 508 MeV~\cite{Suisso:2002jg} 
and 440 MeV~\cite{Yabusaki_2015}; and, the charm constituent 
mass value of 1623 MeV  from~\cite{Suisso:2002jg},
~(see also,  the reference~\cite{Suisso:2002jg} for discussions).
Note that, the constituent quark mass values for all the models 
(A), (C), (E), and (F) are from Ref.~\cite{Suisso:2002jg} in the calculation.
However, we are assuming that the energy scale associated 
with each meson should be the same, which may not be valid. 
We will return to this point later.
 
The static observables considered, i.e., the charge radius and decay constant, 
are shown in  Table~\ref{table1}, where the available experimental data 
for these two quantities
come from Ref.~\cite{Zyla:2020zbs}, and  from \cite{Ablikim:2018jun} 
for the $D^+_s$  weak decay constant.
In addition, the model binding energy shown in the table is given by:
\begin{equation}
\epsilon_M=m_q + m_{\bar{q}} - m_M>0.
\label{binding}
\end{equation}
The pion and kaon appear as strongly bound systems with binding
energies ranging from about 600 to 400 MeV, respectively, for
the best agreement with the charge radius by fitting the decay constants. Both mesons are the
Goldstone bosons of the dynamically broken chiral symmetry, and their
 masses in the chiral limit vanishes according to the
 \linebreak 
  Gell-Mann-Oakes-Renner 
 relation, indicating that these states should form
 strongly  bound quark-antiquark systems with constituent quark degrees of freedom. 
 
\begin{table*}[htb]
\caption{$D^+$ and $D^+_s$ weak decay constants in [MeV], 
and the electromagnetic square radii in [$fm$] for various models.
Experimental data from~\cite{Zyla:2020zbs,Ablikim:2018jun}.
}
\vspace{-0.5cm}
\label{table4}
\begin{small}
\begin{center}
\resizebox{17.0cm}{!}{
\begin{tabular}{|cccccc|}
\hline
\hline 
  Reference	&         ~$f_{D^+}$        &~$f_{D^+_s}$ & $r_{D^+}$ &$r_{D_s^+}$ 
&~$f_{D_s^+}/f_{D^+}$    \\
 			\hline \hline
 			This work      	      & ~144.50  & ~182.70   &  0.505  & 0.377 & 1.26
  \\
 Faessler et al.~\cite{Faessler2003}  & 149.20  & 156.98  &   &   &        1.05 
  \\
  Bashir et al.~\cite{Bashir:2012fs} & ~155.4~    &    205.1  &         &       &  1.32          
  \\ 
Ivanov et al.~\cite{Ivanov:2019nqd}   & 145.7          & 182.2 & &    & 1.25           \\  
 	 Choi~\cite{Choi:2007se}     & ~149.2    &  179.6   &  &                          &   1.20  \\	    	
 				Hwang~\cite{Hwang:2009qz}      & 145.7(6.3) & $189(13)$ 
& $0.406_{+0.014}^{-0.012}$ 
 			&  $0.300^{-0.018}_{+0.023}$ &  1.30(4)  \\
 			
	  			Das et al.~\cite{Das:2015bta}           &   &   & ~0.510 & ~0.465 & \\
 	 	Dhiman \& Dahiya~\cite{Dhiman:2017urn}   & 147.8          & 167.6   &  & & 1.13 \\	
		
		Tang et al.~\cite{Tang:2019gvn}&  295(63) & 313(67) & & &  1.06(32) 
		\\
		& & & & &
	 \\
	 LQCD & & & & &
	 \\	
	 Aubin et al.~\cite{Aubin:2005ar}      & 142(2)(12) & 176(2)(11) & 
 			&  & 1.24(1)(7) \\
 			Follana et al.~\cite{Follana:2007uv}	& 147(3) & 170(2) & &   
&1.16(3)  \\
  Chen et al.~\cite{Chen:2014hva}        & $143.1(1.6)(1.8)$    & 182.9$(0.8)(2.0)$ & &    & 1.28$(3)$ \\ 
  	Carrasco et al.~\cite{Carrasco:2014poa}&146.6(2.6)(0.6)& 174.8(2.8)(1.0)
  	&  & & 1.19(3)(1)
	 \\	
	Can et al. ~\cite{Can2012tx}  &   &  	& 0.371(17) & &  
\\
& & 
  	&0.390(33) & &  
\\
Li \& Wu~\cite{2017EPJA} &  &    	& 0.402(61)   & 0.286(19) &
	\\
	& & 
  	&0.420(82)  &0.354(18) & 
\\		
 & & & & &
 \\
 			PDG~\cite{Zyla:2020zbs}   & 144(3)    &  182(3) &  & 
&  1.26(5)	\\ 
 			Ablikin et al.~\cite{Ablikim:2018jun}   &    & 178.8(2.6) & & &   \\
 			\hline 
 			\hline
 				\end{tabular}
 			}
 		\end{center}
		\end{small}
 		\vspace{2ex}
 	\end{table*}

 The Cutkosky rules~\cite{itzykson2012quantum} 
 applied to the triangle diagram (see Fig.~\ref{fig1}) taking into account our model 
 for the vertex function,
 give that the relevant cuts as function of $m_M$ that have branch points 
 in the regions:
 \begin{equation}
 \mu_M+m_q-m_M>0\quad \text{and} \quad \mu_M+m_{\bar{q}}-m_M>0\, .
 \label{branchp}
 \end{equation}
These branch points are also clear in  the analytic form of Eq.~\eqref{PSILF}
for the wave function.
The minimum value of the position
of the branch point is actually the dominant scale that determines the charge radius. 
From the perspective of the closest value to the continuum, namely
corresponding to the minimum value among $\epsilon_M$ and the branch 
points in Eq.~\eqref{branchp}, we can analise the results for
the parametrisations after fitting the decay constants as given in 
Table~\ref{table1}.

For the pion, one finds for sets 
(A)  and  (B),  $469$ and $680M$~MeV, 
respectively, for the closest branch point to the continuum.
That shows a strongly bound system of constituent quarks, 
and not surprisingly closer values for the two sets than the binding energies. 
 For the kaon,  the sets (C) and (D) present the branching points at 
 310 and 326 MeV, respectively,  
 that are somewhat closer than considering the comparison
 only of the binding energies. 
 
 For the $D^+$, set (E),  one realizes that 
 in agreement with the constraint coming from Eq.~\eqref{branchp},
 the value of $\mu_M\sim m_c$, 
 and the minimum  branch point is actually at $122$~MeV, 
 that is associated with the  charge radius of 0.505~fm, while for the $D^+_s$ 
 we find $225$~MeV, and a radius  of 0.377~fm.
 The   larger value of the $D^+_s$ branch point and the concomitant decreasing 
 of the radius with respect to $D^+$ 
come from the larger value of $f_{D^+_s}$ and $m_s$  in comparison 
to $f_{D^+}$ and $m_d$. Therefore, 
the quarks in $D^+_s$ are in a more compact  configuration  than the 
corresponding ones in $D^+$, 
 and it is expect in general  that
 $r_{ D^+_s} < r_{D^+}$.

 Analogous qualitative explanation of the fact 
 that  $r_{ K^+} < r_{\pi^+}$ should be valid. As we will discuss later on, 
 LQCD results obtained using 
 consistent data sets shows that 
 $r_{ D^+_s} < r_{D^+}$~\cite{Can2012tx,2017EPJA},
 supporting our expectation.

The minimum value of the position of the branch point is actually the dominant
scale that determines the charge radius. From
the perspective of the closest value to the continuum,
namely corresponding to the minimum value among
$\epsilon_M$ and the branch points in Eq.~(\ref{branchp}).

\subsection{Pion and Kaon}

In Table~\ref{table3} we present  our results  for the pion (A) and kaon (C) 
charge radii for the sake of comparison 
with other 
calculations~\cite{Bashir:2012fs,Hutauruk:2016sug,
Maris:2000sk,Faessler2003,Ivanov:2019nqd,Jia:2018ary,daSilva:2012gf,Chen:2019otg}. 
Our collection of results from the literature is  by no means complete, and our intention 
is just to place our model with respect
to a sample that covers continuum approaches to QCD with Euclidean 
Schwinger-Dyson and Bethe-Salpeter,
phenomenological ones with and without 
confinement. In  the table, we also show  the experimental results 
from~\cite{Zyla:2020zbs,Ablikim:2018jun}. 

The results obtained with the Euclidean Schwinger-Dyson and Bethe-Salpeter 
equations with phenomenological kernels that satisfy 
the axial-vector Ward-identities, having dynamical 
chiral symmetry breaking in the SU(3) sector were taken from 
Refs.~\cite{Maris:2000sk} and \cite{Bashir:2012fs}. A modern approach
 along this direction~\cite{Chen:2019otg}  has the 
quark-antiquark interaction  composed by a flavor dependent IR part 
and a flavor independent UV part. 
We also compare with results 
from a relativistic constituent quark model, 
which implements a linear realization of chiral symmetry~\cite{Faessler2003}. 

The solution of the Bethe-Salpeter equation for 
the Nambu and Jona-Lasinio (NJL) model with proper-time
regularization is given in~\cite{Hutauruk:2016sug}.
The results for the covariant confining quark model treated in Euclidean
space were reviewed in Ref.~\cite{Ivanov:2019nqd}.

Calculations performed in LF approaches~\cite{daSilva:2012gf,Jia:2018ary} 
were also presented in Table~\ref{table3}.  In Ref.~\cite{daSilva:2012gf} 
a refined light-front phenomenological model for the pion and kaon elastic form factors,
relying on the use of Pauli-Villars regulators 
in a non-symmetrical form, presents result close to our findings. 
In Ref.~\cite{Jia:2018ary}, a model with color singlet NJL and confining interactions
was studied within basis of light-front quantization.

The pion and  kaon are strongly bound systems of constituent 
quarks in the models presented, and in general they 
are able to provide reasonable reproduction of their decay constants 
and charge radii. This is the main feature learned from Table~\ref{table3}, 
and once the decay constant is reproduced in the strongly bound system 
the charge radius follows~\cite{tarrach1979meson,Gerasimov:1978cp}.

\subsection{$D^+$ and $D^+_s$ mesons}

The comparison of our results and a selection of models from the 
literature~\cite{Faessler2003,Choi:2007se,Hwang:2009qz,Bashir:2012fs,Das:2015bta,Dhiman:2017urn,Ivanov:2019nqd}
is presented 
in  Table~\ref{table4}, together with the outcomes of LQCD 
calculations~\cite{2017EPJA,Follana:2007uv,Aubin:2005ar,Chen:2014hva,Carrasco:2014poa,Can2012tx}. 
 In the table we present 
the charge radii and weak decay constants for $D^+$ and $D^+_s$ mesons, 
as well as the experimental data from
Refs.~\cite{Zyla:2020zbs,Ablikim:2018jun},
and in particular 
the ratio $f_{D_s^+}/f_{D^+}=1.226(31)(2)(3)$~\cite{Zyla:2020zbs}.

The results within Euclidean Schwinger-Dyson and 
Bethe-Salpeter equation framework with phenomenological quark-antiquark interaction kernel that entails
infrared confinement and ultraviolet one-gluon exchange from 
QCD applied to the heavy-light mesons were taken from Ref. \cite{Bashir:2012fs}. 
The calculations within light-cone phenomenological models with confinement  
come from Refs.~\cite{Choi:2007se,Hwang:2009qz,Dhiman:2017urn}. 
A confining potential model in instant form applied to 
describe the heavy-light mesons were used in \cite{Das:2015bta}. 
The results from LQCD for
decay constants were taken from
Refs.~\cite{Aubin:2005ar,Follana:2007uv}, the $D^+$ charge radius comes from~\cite{Can2012tx} and for 
$D^+$ and $D^+_s$ from~\cite{2017EPJA}. In general,
the decay constants are quite close to the experimental values, while for the 
charge radii there is a spread in the theory results.
\begin{table}[hbt]
\begin{center} 
\caption{Pseudoscalar meson static observables with the masses for 
D and D$_s$ mesons from LQCD ensembles (B1),  (C1) used in Refs.~\cite{2017EPJA} 
to compute the EM form factors; and the present model (E,F) parameters: 
$m_c=1623$ MeV,  $m_u=384 $ MeV,   $m_s=508$ MeV,  $\mu_{D^+}=1607$  MeV, and
$\mu_{D^+_s }=1685$ MeV.  The mass and decay constant are given in MeV, while  
the charge radius is given in fm.
}
\bigskip
\begin{tabular}{|cccc|}
\hline
\hline
                  Inputs        &~(B1)           &~(C1) &(E,F )\\
 \hline 
  $m_D^+$          &~1737        &~1824 &~1869 \\
  $m_{D^{+}_{s}}$   &~1801        &~1880 & 1968 \\
\hline
Static observ.  & ~(B1) &~(C1) & (E,F)\\
\hline
 \hspace{0.5cm}$r_{D^+}$~\cite{2017EPJA}          
&~0.402(61)      &~0.420(82) & \\
  $r_{D^+}$                                                                   &~0.347~~~~~       &~0.422~~~~~   & 0.505 \\
  $f_{D^+}$         &205.5~~     &170.3~~  & 144.5\\
& & & \\
  \hspace{0.5cm}$r_{D^+_s}$~\cite{2017EPJA}     
&~0.286(19)  &~0.354(18) & \\
  $r_{D_s^+}$                                  &~0.281~~~~~       &~0.312~~~~~  & 0.377\\
  $f_{D_s^+}$                               &243.3~~    &219.4~~  & 182.7\\
\hline
\hline 
\end{tabular}
\label{table4b}
\end{center}
\end{table}

\begin{table}[hbt]
\begin{center} 
\caption{Flavor decomposition of the charge radii of the 
$D^+$ and $D^+_s$ mesons. Comparison with   LQCD results from 
the linear fit (L) and quadratic fit (Q) extrapolating to the physical 
pion mass from Ref.~\cite{Can2012tx}, and from
the ensembles (B1) and  (C1) used in Refs.~\cite{2017EPJA}
}
\bigskip
\begin{tabular}{|cccc|}
\hline
\hline
 Radius (fm)          &  LQCD~\cite{Can2012tx}   &  
 LQCD~\cite{2017EPJA} &   This work  \\
\hline \hline
  $r_{D^+}$          &   0.371(17) (L)      & 0.402(61) (B1)      & 0.505 \\
                          &  0.390(33) (Q)	&  0.420(82) (C1)	&           \\
                        & & & \\
    $r_{D^+,c}$   &  0.226(24) (L)& 0.17(15) (B1)         &   0.233       \\
                           &  0.272(29) (Q)& 0.20(19) (C1)	&		\\
                         & & &  \\
      $r_{D^+,\bar d}$  & 0.585(57) (L)      &  0.692(61) (B1)     & 0.810 \\
      			 & 0.566(104) (Q)  & 0.718(82) (C1)	&		\\
& & & \\
  $r_{D_s^+}$    &                            &     0.286(19) (B1) & 0.377\\
                           &                          & 0.354(18) (C1) & \\
                           & & & \\
  $r_{D^+_s,c}$     &                           & 0.119(50) (B1) & 0.218 \\
                          &                           &0.222(33) (C1) & \\
                          & & & \\
  $r_{D^+_s,\bar s}$    &                             & 0.461(12) (B1)  & 0.576 \\
                          &                           & 0.545(15) (C1) & \\
\hline
\hline 
\end{tabular}
\label{table4a}
\end{center}
\end{table}
Our results for the charge radii, $r_{D^+} =0.505 $~fm 
and $r_{D^+_s}=0.377$~fm,  are somewhat larger than the ones
computed within LQCD~\cite{Can2012tx,2017EPJA} and the  
models with confinement~\cite{Hwang:2009qz,Das:2015bta}. 
The present model does not have explicit  confinement, as the meson is formed 
as a bound state with about 100 MeV binding energy, and even considering 
that the decay constants were fitted the charge radii, it is not 
strongly constrained. 
Differently from the pion and kaon, the decay constant does not seem  
 to determine definitively the charge radius. We can 
trace back this behavior to the dominant factor $m_c/x$ carried by the valence wave function,
Eq.~\eqref{PSILF}, to the expression for the decay constant, 
Eq.~\eqref{fdecay}, as the charm constituent mass is substantially larger 
than the light quark ones, in other words the heavy quark is close to the 
center of mass of the system, which corresponds to the region that the wave 
function is probed in the weak decay amplitude.
Therefore, naively it is quite reasonable that the light quark in the
BS amplitude is loosely constrained by fitting the decay constant, 
however its contribution to the charge radius is far more important than the 
charm one as one can check e.g.  
in the LQCD calculations~\cite{Can2012tx,2017EPJA}.

The light quark within the heavy meson can explore larger distances where the QCD 
infrared  physics is relevant and from where  
the meson gains weight, i.e., the mass is formed,
and thus, one should expect a correlation of the charge radius and the actual value of mass of the 
charmed mesons found within LQCD.  
 In Table~\ref{table4b}, we compare the charge radius of $D^+$ and $D^+_s$ with the
LQCD results from Refs.~\cite{2017EPJA}. For the $D_s$ meson 
the LQCD radius increases with the mass and indicates that
our result would be compatible, within their uncertainties. 
For the $D$, although there is a slight increase of the radius with the mass,
these LQCD results have large errors to make a firm conclusion from the comparison. 
In the table we also show our calculation for the charge radius
changing the mass of the $D^+$ mesons according to the LQCD values. 
We observe that the tendency of increasing radius by increasing 
the $D^+$ and $D^+_s$ meson masses as found in the LQCD calculation seems
to somewhat reproduced by our model. This feature comes in our model 
due to the decreasing of the binding energy, which leads to the increase of the 
meson size.
This suggests that the quantum mechanical binding mechanism is 
somewhat acting in these heavy-light mesons, even though 
the complexity of the quark confinement mechanism. 

The flavor decomposition of the charge radii of the 
$D^+$ and $D^+_s$ mesons is presented in Table~\ref{table4a}, where we also 
compare our model with   LQCD results from 
the linear fit  and quadratic fit extrapolating to the physical pion mass given 
in Ref.~\cite{Can2012tx}, and from
the ensembles (B1) and  (C1) used in Refs.~\cite{2017EPJA}. 
The flavor contribution to the  $D^+$ and $D^+_s$ charge radius squared are defined as:
\begin{eqnarray}
&&r^2_{D^+,c}=6\tfrac{\partial}{\partial q^2} F_{c \bar{d} c}(q^2)|_{q^2=0} \, ,   \nonumber
\\
&& r^2_{D^+,\bar d}=6\tfrac{\partial}{\partial q^2}F_{\bar{d} c \bar{d}}(q^2)|_{q^2=0}\, ,  \nonumber
\\
&& r^2_{D^+_s,c}=6\tfrac{\partial}{\partial q^2} F_{c \bar{s} c }(q^2)|_{q^2=0}\, ,  \nonumber
\\
&& r^2_{D^+_s,\bar s}=6\tfrac{\partial}{\partial q^2}F_{\bar{s} c \bar{s}}(q^2)|_{q^2=0}\, , 
\end{eqnarray}
 and the relations to the meson charge radius are:
\begin{eqnarray}
&& r^2_{D^+}=\tfrac23 r^2_{D^+,c}+\tfrac13 r^2_{D^+,\bar d}\, , \nonumber
\\
&& r^2_{D^+_s}=\tfrac23 r^2_{D^+_s,c}+\tfrac13 r^2_{D^+_s,\bar s} \, .
\end{eqnarray}
The full charge radii given in Table~\ref{table4a} are largely dominated by
the light quark contribution, and we observe this
property is also shared by the LQCD calculations. 
The heavy quark predominantly is placed close to  the center of mass of the 
heavy-light meson, while the light one is in the region of about 0.6 - 1~fm 
distance from the meson center. 
Amazingly, even being at the confinement 
region the charge radius contribution from the light quark is quite consistent 
with the more recent LQCD results from Refs.~\cite{2017EPJA}, 
although we observe the tendency to overestimate these lattice calculations.

\section{EM Form factors}
\label{sec:EMFF}

\begin{figure}[t]
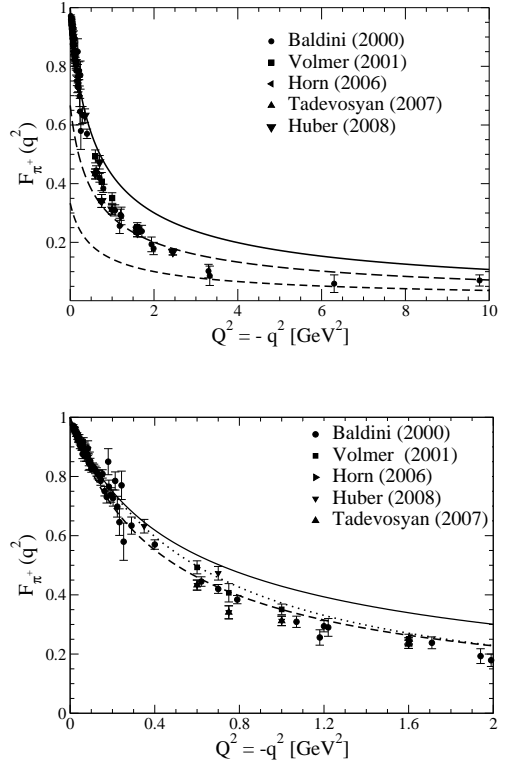

\begin{center}
\epsfig{figure=fig3av1.eps,width=6.50cm,angle=0} \\
\vspace{0.80cm}
\epsfig{figure=fig3bv1.eps,width=6.50cm,angle=0}
\caption{Pion electromagnetic form factor as a function of $q^2<0$. 
Upper panel: 
Flavor decomposition  of $F_{\pi}(q^2)$ for the parameter 
set (A)  (see Table~\ref{table1}). Pion form factor (solid line), 
$u$ contribution - 
$e_uF_{u\bar du}$ (dashed line) and $\bar d$
contribution - $e_{\bar d} F_{\bar d u\bar d}$ 
(short-dashed line). 
Lower panel:  Comparison between the parameter sets (A) (solid line) and (B) 
(dashed line) with the VMD model (dotted line) from Eq. \eqref{VMDP}. 
 Experimental 
data are from Refs.~\cite{baldini2000determination,volmer2001measurement,
horn2006determination,tadevosyan2007determination,huber2008charged}.
}
\label{pi1}
\end{center} 
\end{figure}

\subsection{Pion and Kaon}

The extraction of the pion charge form factor from  the experimental 
cross-section for
 the exclusive pion electroproduction on the proton relies on the dominance of the 
 Sullivan process,  due to the
 pion pole close to the allowed kinematic region for small  $t$'s~\cite{horn2016pion}. 
In principle, it is possible to perform exclusive electroproduction of $K^+$ in
experiments on the proton, to extract the form factor
at higher momentum transfers~\cite{2017APS,horn2016pion}. 
However, in the case of  $K^+$, there is no recent  experimental data for the EM form factor. 
The experimental data obtained at CERN in 1986, come from the most precise measurement 
for the $K^+$ meson~\cite{Amendolia:1986ui} that exist up to the present, and what will
be used in the study of our model.

\begin{figure}[t]
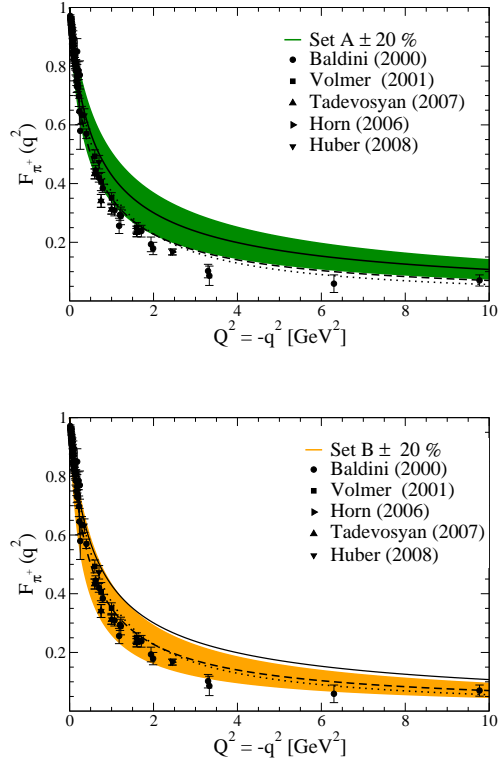

\begin{center}
\epsfig{figure=fig4av1.eps,width=6.50cm ,angle=0} 
\\ \vspace{0.80cm}
\epsfig{figure=fig4bv1.eps,width=6.50cm,angle=0} 
\caption{Pion electromagnetic form factor calculated with variation of the model 
parameters.
Upper panel: band for set (A) with   $\pm$20\%  variation of the model parameters.
Lower panel: band for set (B) with   $\pm$20\%  variation of the model parameters.
In both panels: set (A) (solid line), 
set(B) (dashed line) and VMD (dotted line).
Experimental data are from Refs.
~\cite{baldini2000determination,volmer2001measurement,
horn2006determination,tadevosyan2007determination,huber2008charged}.	
}
\label{pi2}
\end{center} 
\end{figure}
We start the discussion of the light pseudoscalar EM form factors based on
the vector meson dominance~(VMD) model,  formulated 
by Sakurai and others~\cite{Sakurai:1960ju,Sakurai,OConnell:1995nse}. 
The charge form factors of $\pi^+$ and $K^+$ within the  VMD description
are parametrized by the vector mesons masses  $m_\rho$ and $m_\phi$, as: 
\begin{eqnarray}
F_{\pi^+}(q^2)~ & = & ~(e_u+e_{\bar d})\frac{m^2_\rho}{m^2_\rho - q^2} ~ ,
\label{VMDP}
\\
F_{K^+}(q^2)~ & = & ~ e_u \frac{m^2_\rho}{m^2_\rho - q^2} + 
e_{\bar{s}}\frac{m^2_\phi}{m^2_\phi -q^2}~,
\label{VMDK}
\end{eqnarray}
where in the space-like region $Q^2 \equiv -q^2>0$.  
We assume within the SU(3) flavor symmetry that the 
coupling constants of the $\rho$ and $\phi$ are flavor independent,
i.e., $g_{\rho,u}=g_{\rho,d}=g_{\phi,s}$, which implies in the expressions 
written in Eqs.~\eqref{VMDP} and \eqref{VMDK}. From that
 the ratio between the two contributions to the form factor at $Q^2=0$ are 
 the same as the  quark charge ratio, as in our formulation of the
 form factor for $K^+$  due to the charge conservation
 expressed by Eq.~\eqref{ffnorm}. 
 As a matter of fact, this is also verified in our model when the charm and light quark contributions 
are separated in the $D^+$ and $D^+_s$ form factors. 
 
Furthermore, at the level of the VMD, we found that the ratio between the 
quark  contributions to $F_{K^+}(q^2)$  
for~ $Q^2 >> m^2_\phi$ is given by:
\begin{equation}
\frac{e_{\bar s}F_{\bar s u\bar s}}{e_u F_{u\bar s u }} 
{ \frac{m^2_\phi}{m^2_\rho}} \Bigg| _{(q^2 >> m^2_\phi)} 
\to ~ \frac12 \frac{m^2_\phi}{m^2_\rho}=0.86\, ,
\end{equation}
where  the experimental meson masses are used~\cite{Zyla:2020zbs}. 
For comparison, the present model 
gives $\simeq 0.56$ (at 10 GeV$^2$) with parameters from set (C) 
(see Tables~\ref{table1} and \ref{tabpiK}).

\begin{figure}[t]
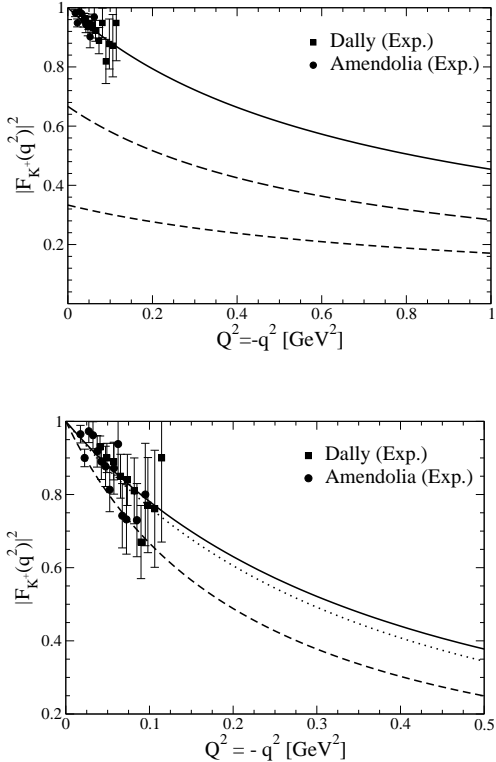

\begin{center}
\epsfig{figure=fig5av1.eps,width=6.50cm,angle=0} 
\\ \vspace{0.80cm}
\epsfig{figure=fig5bv1.eps,width=6.50cm,angle=0}  
\caption{Kaon electromagnetic form factor as a function $q^2<0$. 
 Upper panel: kaon form factor (full line), $u$ contribution - $e_uF_{u\bar su}$ 
 (dashed line) and $\bar s$ 
contribution - $e_{\bar{s}} F_{\bar{s}u\bar{s}}$
(short-dashed line),
computed with the parameter set (C) (see Table~\ref{table1}).
Lower panel:  $|F_{K^+}|^2$, comparison between the results 
from parameter sets (C)
(full line), (D) (dashed line)  and
the VMD model  from Eq.~\eqref{VMDK} (dotted line).
Experimental 
data  from~\cite{Amendolia:1986ui,Dally:1980dj}. }
\label{kaon1}
\end{center} 
\end{figure}

\begin{figure}[htb]
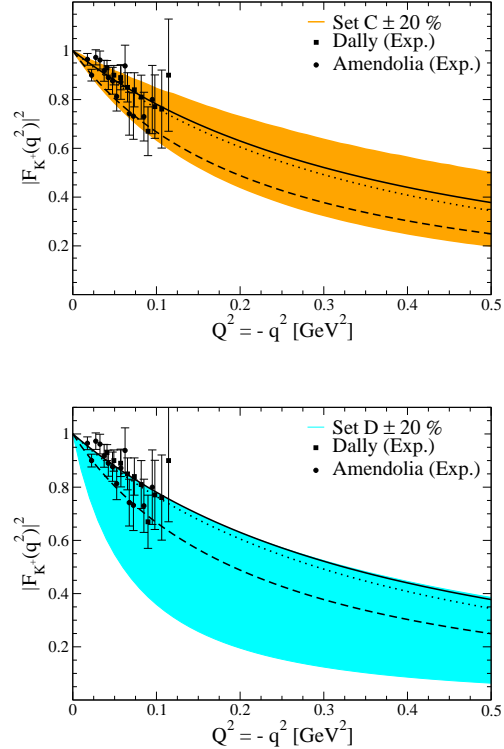

\begin{center}
\epsfig{figure=fig6av1.eps,width=6.50cm,angle=0}
\\ \vspace{0.80cm}
\epsfig{figure=fig6bv1.eps,width=6.50cm,angle=0}
\caption{
 Kaon electromagnetic form factor square with  variation of the model parameters.   
Upper panel: band for set (C) with   $\pm$20\%  variation of the model parameters.
Lower panel: band for set (D) with   $\pm$20\%  variation of the model parameters.
In both panels: set (C) (solid line), set(D) (dashed line) and VMD (dotted line).
Experimental data from Refs~\cite{Amendolia:1986ui,Dally:1980dj}.		
}
\label{kaon2}
\end{center} 
\end{figure}

In Fig.~\ref{pi1}, we present the  results for the pion EM form factor 
obtained with the parameter sets (A), (B) (see Table~\ref{table1}) and the VMD model in Eq.~\eqref{VMDP}. 
In both panels the 
experimental data  from \cite{baldini2000determination,volmer2001measurement,horn2006determination,
tadevosyan2007determination,huber2008charged}
are also shown. 
In the upper panel of the figure, the individual quark contributions are
shown up to 10~GeV$^2$ for set (A), and
the contributions from the 
$u$ and $\bar d$ quarks are in the
ratio 2:1, as it should be from the  model with SU(2) flavor symmetry. 

The comparison between the results for sets (A), (B)  and the VMD model is presented
in the lower panel of Fig.~\ref{pi1}. %
Note that the set (A) reproduces well the experimental values of $\pi^+$ charge radius 
and decay constant, while set (B) performs better at large momentum transfers,
but the charge radius is overvalued  (see Table~\ref{table1}).
However, both sets have form factors with the same analytical behavior  
at large momentum transfer region, and only 
their normalizations differ by a couple of tenths of percent. This suggests
a limitation in the chosen form of the  vertex function to 
describe the pion charge distribution. Indeed, models with running quark
masses and incorporating the asymptotic QCD counting rules 
are known to perform better with respect to the 
experimental data 
(see e.g. \cite{Maris:2000sk,Bashir:2012fs,Mello:2017mor}).

In order to overcome the limitation of the model,  
we allowed a $\pm$20\% parameter variation in sets (A) and (B),
keeping the pion mass fixed in Fig.~\ref{pi2}. 
This changes  both the decay constant and charge radius, while 
it allows that results with the variation of set (A) becomes 
much closer to the experimental data, set (B) produces a 
band englobing almost all the experimental data. 
In the chiral limit, where the pion becomes massless the changes in the 
constituent quark masses and mass scale parameter by a 
factor $\lambda$ will give for the new form factor, $f'_\pi(q^2) $, 
the scaling relation:
\begin{equation}\label{scaling}
f'_\pi(q^2)=f_\pi(\lambda^2q^2)\, ,
\end{equation}
and it represents reasonably  well the band that one sees in the figure. 

A last comment on the pion form factor is in order. 
The  pion and heavy meson  energy scales can be different, as we have
pointed out earlier. The pion is a strongly bound system of the 
constituent quarks, and therefore  should be associated 
with a somewhat larger energy scale  with respect to the heavy meson ones.
This explains why the pion form factor at large momentum transfers is better 
reproduced with model B having lighter u and d quarks, in contrast with model A,
where the quark mass is appropriate for the heavy mesons.

We present the results for the $K^+$ form factor in Figs.~\ref{kaon1} and~\ref{kaon2} 
obtained with the two sets of parameters (C)  
and (D) (see Table~\ref{table1}). We also compare the results with the VMD model and the 
experimental data from~\cite{Amendolia:1986ui,Dally:1980dj}. We remind that the set (C) 
presents  a charge radius and weak decay constant in agreement with the experimental values.
The parameter set (D) reproduces the decay constant but
has about 20\% difference with the experimental value of the kaon charge radius. 
In the upper panel of Fig.~\ref{kaon1},  we show  
the results obtained with the set C, 
for the full  and the $u$ and $\bar s$ quark contributions for the  kaon  form factor. 
In the figure, it is possible to see, that the SU(3) flavor symmetry is
slightly broken, as the ratio
$e_{\bar s} F_{\bar su\bar s}/(e_uF_{u\bar su})>1/2$ for $Q^2\neq0$. 
The same feature is also found in the NJL model~\cite{Hutauruk:2016sug}.

In the lower panel of Fig.~\ref{kaon1} we compare the 
 results for the kaon form factor in the low momentum transfer region up to 
 0.5~GeV$^2$ obtained with  set (C), (D) 
and the VMD model
together with the experimental data~\cite{Amendolia:1986ui,Dally:1980dj}. 

One can see that this data cannot  unambiguously resolve between sets (C) and (D), 
with both somewhat consistent with the experimental results.
From Table~\ref{table1}, we have that set (C) reproduces the experimental 
kaon charge radius
and it is also consistent with the VMD form factor, therefore taking that into 
account  we could say that it presents an overall better 
consistency with the data.
 
\vspace{0.45cm}
\begin{figure} [htb]
\begin{center}
\mbox{
\epsfig{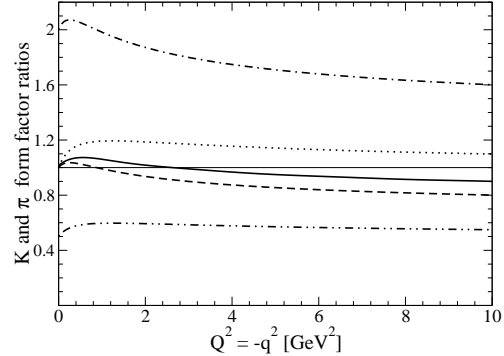}
} 
\caption{The electromagnetic form factor ratios for the pion and kaon 
using sets (A) and (C), respectively. 
Ratios of flavor contributions to the pion and kaon form factors: 
$F_{K^+}/F_{\pi^+}$ (solid line); 
$e_uF_{u\bar s u}/(e_{\bar d}F_{\bar d u\bar d})$ (dot-dashed line),
$ F_{u\bar s u}/F_{u\bar d u}$ (dashed line);
$ F_{\bar s u\bar s}/F_{\bar d u\bar d }$ (dotted line);
$e_{\bar s} F_{\bar s u\bar s}/(e_u F_{u\bar s u })$ (dot-dot-dashed line).
A thin solid line is  the reference for the SU(3) flavor symmetry. 
}
\label{piK}
\end{center} 
\end{figure}

In the upper panel of Fig.~\ref{kaon2}, we examine the effects of
a $\pm$20\% variation 
in the model parameters on the kaon EM form factor using set (C). The 
band now  englobes
the experimental data for the form factor. 
Although, the validity of an 
analogous scaling property 
as verified for the pion in Eq.~\eqref{scaling} is questionable, as the 
constituent quark masses,  
mass scale parameter and binding energy are large enough, a similar band 
is found as in the pion case.
However, the 20\% change in the parameters of set (D) keeping the kaon
mass fixed as presented in the 
lower panel of the figure, shows a quite large  band.  This is due to the fact that 
decreasing the masses in set (D) by 20\% 
makes the binding energy drops from 166 MeV (Table~\ref{table1}) to only 34 MeV, 
producing a large increase in the
radius, and consequently the wide band observed in the figure.

The  breaking of the  SU(3) flavor symmetry is analyzed using the sets 
(A) and (C)  in Fig.~\ref{piK}, 
where we show several ratios of form factors to 
evidence such effect in the space-like region. The ratio of the 
kaon and pion form factors goes above unity at low momentum reflecting
the  more compact charge distribution of the kaon with respect to the pion
one, while at large momentum transfers we observe
that the ratio stays below the unity. This can be traced back to the fact that set (A) 
produces a pion form factor above the experimental data at large momentum transfers, 
as well as above the
VMD results (see Fig.~\ref{pi2}), while the kaon form factor for set (C)
agrees with the VMD one (see  the lower panel of Fig.\ref{kaon1}).
The VMD model from Eqs.~\eqref{VMDP} and\eqref{VMDK} gives 
 the ratio at large momentum transfers:
\begin{eqnarray}
\frac{F_{K^+}(q^2)}{F_{\pi^+}(q^2)}  \to ~ e_u  + 
e_{\bar{s}}\frac{m^2_\phi}{m^2_\rho}=1.24~,
\label{eqvmd1}
\end{eqnarray}
while QCD for $Q^2>>\Lambda^2_{QCD}$ predicts that~\cite{LepPLB1979}:
\begin{eqnarray}
\frac{F_{K^+}(q^2)}{F_{\pi^+}(q^2)}  \to \frac{f^2_K}{f^2_\pi}=1.42 \pm 0.03\, ,
\label{eqvmd2}
\end{eqnarray}
where we have used the experimental value for the decay constant ratio~\cite{Zyla:2020zbs}.
If we had chosen set (B) for the pion with set (C) for the kaon to compute 
 $F_{K^+}(q^2)/F_{\pi^+}(q^2)$,  we would have had a value somewhat larger than unity, 
 towards the QCD results, however the model doesn't have 
 the perturbative QCD dynamics built in, and we do not expect 
 to reproduce Eq.~\eqref{eqvmd2} in the asymptotic momentum region. 
  Indeed, emphasizing that our model is to be applicable at low momentum transfers, we computed for $Q^2=200$~GeV$^2$  the product $Q^2F_\pi(Q^2)$,  which is already saturated around a value of
 1.96~GeV$^2$ for set A and 1~GeV$^2$ for  set B  from Table~\ref{table1}, while the asymptotic QCD formula 
 $Q^2F_\pi(Q^2)\to 16 \pi\alpha_s(Q^2) f^2_\pi$~\cite{LepPLB1979} for $Q^2=200~\text{GeV}^2$ gives $\sim 0.1$~GeV$^2$ with $\alpha_s(Q^2)$ from~\cite{Zyla:2020zbs}.

In addition, in Fig.~\ref{piK} the ratios of flavor contributions for the pion and 
kaon form factors are 
shown. The ratio of the $u$ quark contribution to the kaon and pion
form factors, denoted  by $F_{u\bar s u}/F_{u\bar d u}$
follows a similar trend as the ratio of the full form factor, as well as
$e_uF_{u\bar s u}/(e_{\bar d}F_{\bar d u\bar d})$ apart the factor 2 from the charge ratio,
as for the pion $F_{u\bar d u}=F_{\bar d u\bar d}$.
At small
momentum transfers, we observe that the configuration where photon is
absorbed by the $u$ quark 
in the pion is somewhat  larger than in the kaon.
The ratio of $\bar s$ contribution in the kaon to the $\bar d$ one in the pion given by
$ F_{\bar s u\bar s}/F_{\bar d u\bar d }$ presents a dependence  on 
$Q^2$, similar as we have discussed, i.e., it reflects the
more compact configuration of the kaon with respect to the pion. We should say,
that despite the large momentum behavior of the pion form factor for 
set (A) overestimate the experimental data, its prediction in lower momentum transfer 
region should be reasonable. 
As shown in the figure, the ratio between the $\bar s$ and $u$ contributions 
to the kaon form factor, namely 
$e_{\bar s} F_{\bar s u\bar s}/(e_uF_{u\bar s u })$,
clearly is above the flavor symmetric value of 1/2 for 
larger momentum transfers as 10~GeV$^2$. Such an effect is quite visible in 
Fig.~\ref{kaon1}, when looking  to the individual  flavor contribution to the
kaon charge form factor.

\begin{table}[htb]
	\begin{center}
		\caption{Partial ratios for the electromagnetic
			form factors of pion (A) and  kaon (C) at 10~GeV$^2$,
			compared with the 
		NJL  model  from Ref.~\cite{Hutauruk:2016sug}. 
		}		
		\label{tabpiK}
		\begin{tabular}{|l|l|l|l|}
			\hline
			\hline
			Model &~${F_{u\bar s u}\over F_{u\bar d u}}$        
			& ${F_{\bar s u \bar s}\over F_{\bar d u \bar d}}$  
&  ${e_{\bar s} F_{\bar s u \bar s}\over e_u F_{u\bar s u}}$    
\\ 
			\hline 
			This work                                   &~0.80        &~1.10    &  0.69 \\	
			NJL~\cite{Hutauruk:2016sug}                 &~0.36      & ~2.74     &  0.56  \\
			\hline 
			\hline
		\end{tabular}
	        \end{center}
\end{table}

\begin{table}[thb]
\begin{center}
\caption{Ratio of the electromagnetic
form factors for pion (A)  and kaon (C), compared 
with  calculations from~\cite{Hutauruk:2016sug,Bakulev:2001pa,Shi:2014uwa} and 
experimental time-like data~\cite{Pedlar:2005sj,Seth:2012nn}.
}
\label{tabpiK2}
\begin{tabular}{|c|c|c|}
\hline 
\hline 
 Reference & $Q^2$~[(GeV/c)$^2$] & ~$F_{\pi^+}/F_{K^+}$     \\ 
\hline 
This work     &   ~10.0            & ~1.10      \\ 
              &   ~13.48           & ~1.14       \\
              &   ~14.2            & ~1.13        \\
              &   ~17.4            & ~1.16         \\
Hutauruk et al.~\cite{Hutauruk:2016sug}  & ~10.0     & ~0.87   \\
Bakulev et al.~\cite{Bakulev:2001pa}     & ~13.48    & ~0.53    \\
Shi et al.~\cite{Shi:2014uwa}            & ~17.4     & ~0.81  \\
\hline
Pedlar et al.~\cite{Pedlar:2005sj}       &~13.48      & ~1.19(17) \\
Seth et al.~\cite{Seth:2012nn}           &  ~14.2     & ~1.21(3)  \\
Seth et al.~\cite{Seth:2012nn}           &  ~17.4     & ~1.09(4)  \\
\hline 
\hline
\end{tabular} 
\end{center}
\end{table}

In Table~\ref{tabpiK}, the ratios of flavor form factors of the pion and kaon 
for the sets (A) and (C), respectively, and presented in Fig.~\ref{piK}  are
compared with the results from the NJL  model obtained in Ref.~\cite{Hutauruk:2016sug} 
at the particular value of $Q^2=10$~GeV$^2$.
The results in the table illustrate once more the SU(3) flavor symmetry 
breaking  through the flavor form factor ratios.
Our results show the deviation of ${F_{u\bar s u}/F_{u\bar d u}}$ and 
${F_{\bar s u \bar s}/F_{\bar d u \bar d}}$ 
from unity, which are considerably smaller than in the NJL model reflecting that 
set (A) overestimates the pion form factor at large momentum.  
The results of the NJL model from Ref.~\cite{Hutauruk:2016sug} show a larger 
symmetry breaking for these ratios. 
The ratio between the  $\bar s$ and $u$ flavor contributions  to the kaon form factor
(fourth column in Table~\ref{tabpiK}) are consistent, even taking into account
that quark current  is dressed in Ref.~\cite{Hutauruk:2016sug}, 
and in addition it shows that our Bethe-Salpeter model has  a small, namely 
about 10\%,
flavor symmetry breaking for the up and strange quarks within 
the kaon. One important difference to be pointed out 
is that the NJL model as studied in Ref.~\cite{Hutauruk:2016sug}, has 
infrared and ultraviolet scales, while our model
is less flexible, having just one mass scale, which makes the present model more 
robust against flavor symmetry breaking among the quark contributions
to the kaon form factor.

From the Phragm\'en-Lindel\"of’s theorem it follows that at large momentum 
transfers the asymptotic behavior
of the form factors in the space-like and time-like regions must be the 
same~\cite{Bilenky:1993cd,Denig:2012by}. Therefore,
in principle, we could compare our results for space like form factors 
at large momentum transfers 
with the experimental time-like ones~\cite{Pedlar:2005sj,Seth:2012nn}.
In Table~\ref{tabpiK2}, we present the ratios of $F_{\pi^+}/F_{K^+}$ 
at some large $Q^2$ values for our model calculated with sets (A) and (C), 
compared to results
from other models~\cite{Bakulev:2001pa,Hutauruk:2016sug,Shi:2014uwa}. 
In the flavor SU(3) symmetry limit, all the ratios presented in the table 
should be unity, 
which is clearly not the case.
Furthermore, QCD predicts that for $Q^2>>\Lambda_{QCD}$ the ratio between the form factors 
approaches
$f^2_\pi/f^2_K \simeq 0.70$~\cite{LepPLB1979}, which suggests by comparing this ratio with 
the time-like experimental data,  that
$Q^2\sim $17~GeV$^2$ is still not in  the asymptotic region. In particular our results for the ratio using 
set (A) and (C), for the pion and kaon  respectively, overestimate the ratio, 
and if we had used instead set (B) for the pion we would have found
a value below the unity, as the other models.
We emphasize,  that the parameter set (A)  for the pion was chosen to analyze the pion, 
as it gives a better representation of the form factor at low momentum transfers, 
and in addition  larger light quark masses  
of this set
 avoid to unbind   the $D^+$ and $D^+_s$, as we have already  shown when discussing 
 the static observables of these heavy-light mesons.

\begin{figure}[htb]
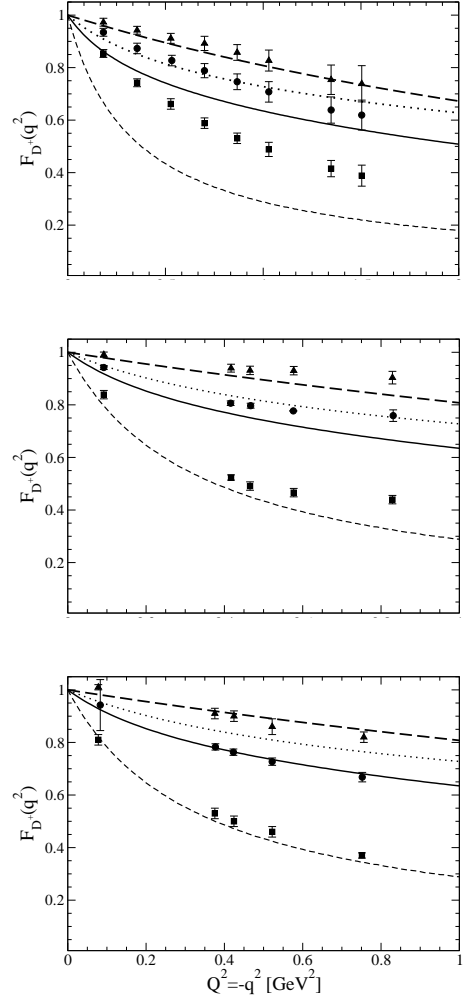

\begin{center}
\epsfig{figure=fig8av1,width=6.cm ,angle=0} 
\\ \vspace{0.20cm}
\epsfig{figure=fig8bv1.eps,width=6.cm ,angle=0} 
\\ \vspace{0.20cm}
\epsfig{figure=fig8cv1.eps,width=6.cm,angle=0}  
\caption{$D^+$  electromagnetic 
form factor with the corresponding quark contributions and comparison with
lattice calculations. Our results: $D^+$ form 
factor  (full line), $\bar d$ contribution - $e_{\bar d}F_{\bar d c \bar d}$  
(short-dashed line) and $c$ contribution -  
$e_cF_{c \bar d c}$ (dashed line). VMD (dotted line) from Eq.~\eqref{VMDD}.
LQCD results for 
$D^+$ form factor (circles),  $\bar d$ contribution (squares) and $c$  contribution
(triangles).
 Upper panel: comparison with LQCD results from Ref.~\cite{Can2012tx}. 
 Middle panel: comparison with LQCD results from 
 ensemble (B1)~\cite{2017EPJA}. 
 Lower panel: comparison with LQCD results 
 from ensemble (C1)~\cite{2017EPJA}.
}
\label{Dlattice}
\end{center}
\end{figure}

\begin{figure}[htb]
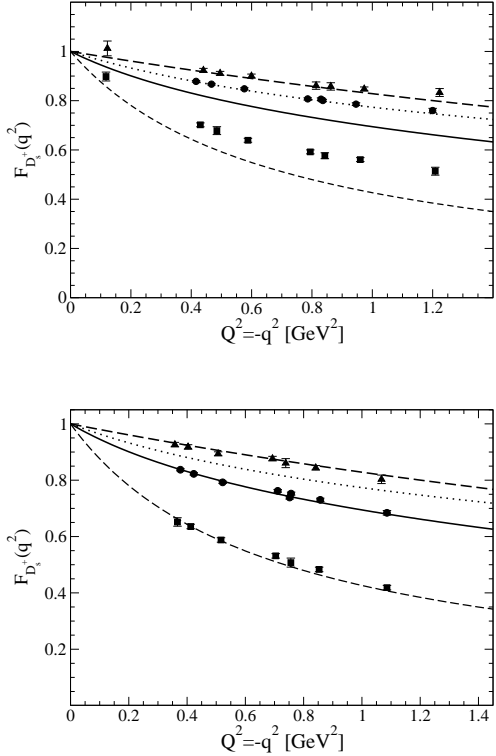

\begin{center}
\epsfig{figure=fig9av1.eps,width=6.5cm ,angle=0} 
\\ \vspace{0.80cm}
\epsfig{figure=fig9bv1.eps,width=6.5cm,angle=0}  
\caption{$D^+_s$  electromagnetic 
form factors with the corresponding quark contributions and comparison 
with lattice calculations. Our results: $D^+_s$ form factor  (full line), 
$\bar d$ contribution (short-dashed line) and $c$ contribution (dashed line), 
VMD (dotted line) from Eq.~\eqref{VMDD}. LQCD results for
$D^+_s$ form factor (circles),
$\bar s$ contribution (squares) and $c$  
contribution  (triangles).  Upper panel: comparison with LQCD results from
 ensemble (B1)~\cite{2017EPJA}. Lower panel:
 comparison with LQCD results from
 ensemble (C1)~\cite{2017EPJA}.
}
\label{Dslattice}
\end{center}
\end{figure}

\subsection{$D^+$ and $D^+_s$ mesons}

We study in what follows the $D^+$ and $D^+_s$ EM form factors and their 
flavor decomposition.
In particular, we illustrate quantitatively the manifestation of the SU(4) 
flavor symmetry breaking  on the
contributions of each quark to the form factor.  We have used 
the parameter sets (E) and (F) given
in Table~\ref{table1}, which fit the  $D^+$ and $D^+_s$ decay constants, respectively. 
The masses of these heavy-light mesons are the experimental ones and given in the table.
The comparison 
will be made with LQCD results of the form factors for
$D^+$~\cite{Can2012tx,2017EPJA} for 
$D^+_s$~\cite{2017EPJA},
even though  the physical masses of these mesons were not achieved in these calculations. 
This has an impact on the charge radius, as it was verified
in our discussion of Table~\ref{table4b}.
However taking into account the quoted errors of the lattice calculations  
our results are quite consistent with ensemble C1 used 
in Refs.~\cite{2017EPJA}.
Such a trend will be confirmed by comparing  the form factors and 
the corresponding quark contributions with such LQCD results.

\begin{figure}
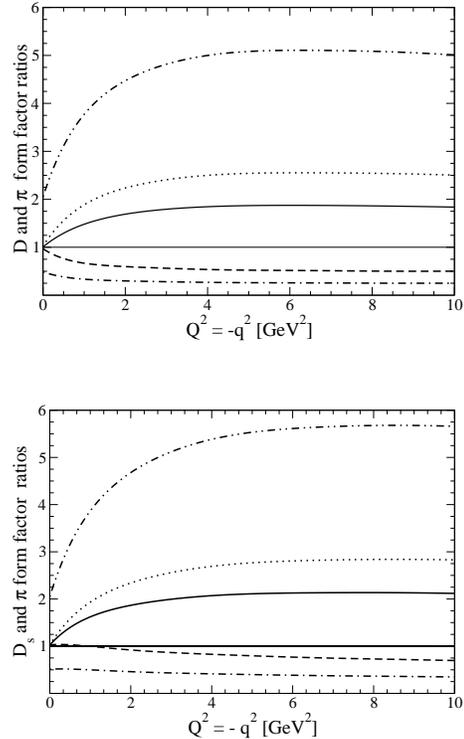
  
\begin{center}
\epsfig{figure=fig10av1.eps,width=6.cm,angle=0}  
\\ \vspace{0.80cm}
\epsfig{figure=fig10bv1.eps,width=6.cm,angle=0}
\caption{The electromagnetic form factor ratios 
for the pion (A) and $D$ mesons. 
Upper panel: $F_{D^+}/F_{\pi^+}$ (solid line); 
$F_{\bar d c\bar d}/ F_{\bar d u\bar d}$ (dashed line);
$e_{\bar d}F_{\bar d c\bar d}/ (e_uF_{ u\bar du})$ (dot-dashed line);
$F_{c\bar dc}/ F_{u\bar d u}$ (dotted line); 
$e_cF_{c\bar dc}/(e_{\bar d} F_{\bar d u \bar d})$ (dot-dot-dashed line).
Lower panel: $F_{D^+_s}/F_{\pi^+}$ (solid line); 
$F_{\bar s c\bar s}/ F_{\bar d u\bar d}$ (dashed line);
$e_{\bar s}F_{\bar s c\bar s}/ (e_uF_{ u\bar du})$ (dot-dashed line);
$F_{c\bar sc}/ F_{u\bar d u}$ (dotted line);
$e_cF_{c\bar sc}/(e_{\bar d} F_{\bar d u \bar d})$ (dot-dot-dashed line).
The thin solid line  is the reference for the  SU(4) flavor symmetry.
}
\label{piD}
\end{center} 
\end{figure}

To have a phenomenological handle on the computed form factors, we exploit the VMD 
applied for $D^+$ and $D^+_s$:
\begin{eqnarray}
 F_{D^+}(q^2)&  = & \frac{2}{3} 
 \frac{m^2_{J/\psi}}{m^2_{J/\psi}- q^2} 
  + 
 \frac{1}{3} 
  \frac{m^2_\rho}{m^2_\rho- q^2} 
 \,,  \label{VMDD}
\\
 F_{D^+_s}(q^2)&  = & \frac{2}{3} 
\frac{m^2_{J/\psi}}{m^2_{J/\psi}- q^2} + 
 \frac{1}{3} 
 \frac{m^2_\phi}{m^2_\phi- q^2}\,. 
 \label{VMDDS}
 \end{eqnarray}
 The expressions for the form factors based on the VMD where the masses 
 of $J/\Psi$, 
 $\rho$ and $\phi$ determines the closest poles to the space-like momentum region 
 in the photon-absorption 
 amplitude. The values of the vector meson masses  come  from~\cite{Zyla:2020zbs}.
The VMD model form factor gives for $D^+$  a charge radius  of
$r_{D^+}=0.381$~fm and for $D^+_s$  it gives  $r_{D_s^+}=0.302$~fm. 
These values are somewhat close to the present model and also to  the 
lattice results (see the Table \ref{table4}).

The results for the EM  form factors of the $D^+$ and $D^+_s$ mesons 
are presented in Figs.~\ref{Dlattice} and \ref{Dslattice}, respectively.
In addition the flavor decomposition and the comparison with
LQCD results from~\cite{Can2012tx} and~\cite{2017EPJA} are shown.
 The three panels of Fig.~\ref{Dlattice} are dedicated to the $D^+$  form factor, 
 where our calculations
  with parameter set (E) given in Table~\ref{table1}, and the corresponding quark 
  contributions, namely
   $F_{\bar d c\bar d}$ and $F_{c\bar d c}$, are compared with 
LQCD results and the VMD model from Eq.~\eqref{VMDD}. 
In the upper panel, the LQCD  form factors~\cite{Can2012tx} are shown 
together with the VMD, and set (E) models.
As anticipated in the analysis of the charge radii in Table~\ref{table4a} we 
observe consistently that our model form factors 
are below the results from~\cite{Can2012tx} up to 2~GeV$^2$.

 The heavy quark contribution to the $D^+$ form factor shown in Fig.~\ref{Dlattice} decreases 
slowly as it sits at the center of mass of the meson, while the light quark 
form a  "halo" charge distribution around the heavy 
quark, exploring the confinement region. Despite of that the behavior of 
$F_{\bar d c\bar d}$  with $Q^2$ is smooth 
making useful  our model to represent the  heavy-light meson charge distribution. 
The LQCD calculations from Ref.~\cite{2017EPJA} were 
done with two ensembles, namely (B1) and (C1), with the last one providing the 
closest value of the $D^+$ meson mass 
to the experimental value (see Table~\ref{table4b}). These LQCD results are  
compared to our model, which are 
shown in the  figure middle and lower panels. Up to the errors of the form factor 
computed with ensemble C1, we found agreement with our model, which does not happen 
with the VMD form factor. 
The light quark
"halo" charge form factor from LQCD ensemble C1 is also reproduced, as well as the 
charm contribution to the $D^+$
form factor.

In  Fig.~\ref{Dslattice}, we show results for the $D^+_s$ electromagnetic form factor 
and the corresponding 
flavor decomposition for our model obtained with the parameter set (F). The comparison 
with VMD model and LQCD 
calculations from Ref.~\cite{2017EPJA}  with ensembles B1 (upper panel) and
C1 (lower panel) is alo shown.
The form factor $F_{D^+_s}$,  and flavor contributions $F_{c\bar sc}$ and 
$F_{\bar s c\bar s}$ underestimate the
 results from  the LQCD ensembles B1, and the VMD model,  are given 
 in the upper panel of the figure. 
 This result  is expected as the computed radii are larger than the ones 
 obtained with ensemble B1 
 (see Table~\ref{table4a}). The ${D^+_s}$ form factor and its flavor content 
 obtained with our model are in 
 agreement with the outcomes from the lattice ensemble C1 as shown in the lower 
 panel of the figure, a 
 finding consistent with the one observed for $D^+$ presented in the  
 lower panel of Fig.~\ref{Dlattice}. 
  The consistency found for both $D^+$ and $D^+_s$ form factors and flavor
  content with 
 the LQCD results from  ensemble C1, suggests that the IR  physics embedded 
 within our parametrization and 
 constituent masses reflect consistently the QCD long-distance effects in these
 heavy-light mesons. Important to 
 observe that our choice of parametrization (A) for the pion and (C) for the kaon,
 describe both their charge radii 
 and decay constants, quantities carrying the QCD nonperturbative IR physics.

\begin{figure}[thb]
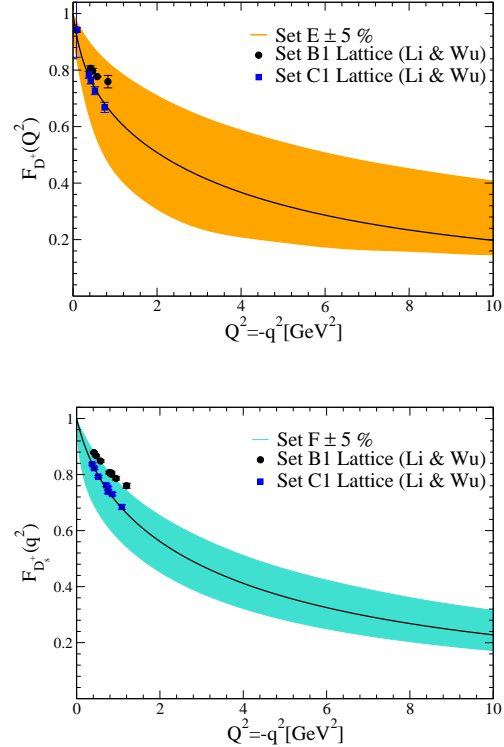

\begin{center}
\epsfig{figure=fig11av1.eps,width=6.50cm,angle=0}  
\\ \vspace{0.80cm}
\epsfig{figure=fig11bv1.eps,width=6.50cm,angle=0}   
\caption{$D^+$ and $D^+_s$ electromagnetic form factors,  
together with bands for parameter variations.
Upper panel: $D^+$ form factor band for set (E) with   $\pm$5\%  variation of
the model parameters.
Lower panel: $D^+_s$ form factor band for set (F) with   $\pm$5\%  variation 
of the model parameters. For reference the
LQCD results from ensembles B1 (circles)  and C1 
(squares)~\cite{2017EPJA} 
are shown.
}
\label{DDs2}
\end{center}
\end{figure}

\begin{figure}[thb]
\vspace{0.2cm}
\begin{center}
\epsfig{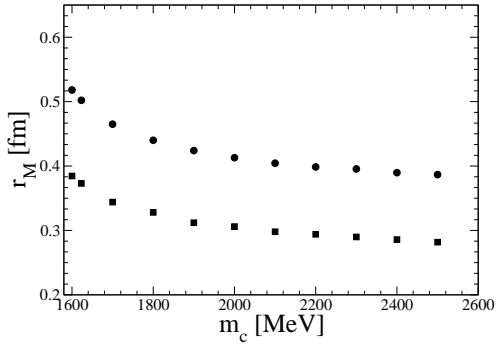}
\caption{
Charge radii of  $D^+$ and $D^+_s$  mesons as a
 function of  the  charm quark mass, $m_c$, having all other parameters 
 fixed to set (E) for
$D^+$ (circles)  and to set (F) for $D^+_s$ (squares) .
}
\label{Dradius1}
\end{center} 
\end{figure}
 
 In the following we discuss  the manifestation of the  SU(4) symmetry breaking  
 in $D^+$ and $D^+_s$ form factors. 
For that aim in Fig.~\ref{piD}, we present several ratios of the electromagnetic 
form factors of the $D$ mesons and their flavor components with the pion ones from model (A).
In the upper panel, we present the  ratio $F_{D^+}/F_{\pi^+}$,  which tends to a 
value close to 2 for large momentum
transfers. This ratio flattens above $Q^2=-q^2\gtrsim 3$ GeV$^2$ that indicates 
the momentum region where the difference in the constituent quark masses and values 
of the regulator scales $\mu_M$ are somewhat irrelevant for the dependence 
on $q^2$, while the value of the ratio $ \sim 2$ seems to be more particular 
to our model. However, we should convey that the normalizations of the form factors 
are essentially determined at low momentum scales which should be more constrained 
with the fitting of the decay constant and charge radius. On the other hand, 
taking into account the ratio of decay constants squared the present value is 
about half of the value that it should be for $F_{D^+_s}/F_{\pi^+}$ (lower panel), 
while for $F_{D^+}/F_{\pi^+}\sim 2$ is consistent with that ratio.

The ratios of the $\bar d$ contribution to the $D^+$ and $D^+_s$ 
to pion form factor, namely
$F_{\bar d c\bar d}/ F_{\bar d u\bar d}$ (upper panel) and 
$F_{\bar s c\bar s}/ F_{\bar d u\bar d}$ (lower panel), 
respectively, exemplify two aspects of the SU(4) flavor symmetry breaking, 
a strong one in the distribution of the $\bar d$ charge in the $D^+$, 
which is extended with respect to the one in the pion, while the effect 
is somewhat weaker in the
$ D^+_s$, where the $\bar s$ is heavier  than $\bar d$ and its charge 
distribution in $D^+_s$ is similar to the pion one. This last observation 
is corroborated by the similar radii $r_{D^+_s,\bar s}$ 
and $r_\pi$ (see Tables \ref{table1} and \ref{table4a}). 
The ratios $e_{\bar d}F_{\bar d c\bar d}/ (e_uF_{ u\bar du})$ (upper panel) 
and $e_{\bar s}F_{\bar s c\bar s}/ (e_uF_{ u\bar du})$ (lower panel) 
just reflect what we have already discussed. The heavy quark contribution 
to the $D^+$ and $D^+_s$ form factors and the associated evidence of SU(4) flavor 
symmetry breaking is clearly seen in the ratios
$F_{c\bar dc}/ F_{u\bar d u}$ (upper panel)  and $F_{c\bar sc}/ F_{u\bar d u}$ (lower panel), 
which saturate above $\sim 3$~GeV$^2$, attaining a ratio about 3, in connection 
with the strong localization of the heavy quark at the meson center of mass. 
Such large  ratio would be even increased if we had chosen a pion model 
that fits the form factor like model (B) at large momentum, and the  flavor 
symmetry breaking would be even enhanced. 
These characteristics are  visible in the 
ratios $e_cF_{c\bar dc}/(e_{\bar d} F_{\bar d u \bar d})$ (upper panel) 
and $e_cF_{c\bar sc}/(e_{\bar d} F_{\bar d u \bar d})$ (lower panel).

In Fig.~\ref{DDs2}, we show the effect of the variation  
of 5\% in the parameters
 for both the form factors of $D^+$ (upper panel) and 
 $D^+_s$ (lower panel). For reference, we include in the figure
 the LQCD results from~\cite{2017EPJA}. 
 The band width is larger for the case of $D^+$ because 
 it has a smaller biding energy, making the form factor much more sensitive to  
 changes in the quark  masses and regulator mass $(\mu_M)$, while 
 for the $D^+_s$ one observes a smaller band width. 
 In both cases such parameter change is enough to englobe the 
 LQCD results from ensembles B1 and C1 of Refs.~\cite{2017EPJA}, 
 and for $D^+$ also the calculations from Ref.~\cite{Can2012tx} not shown in the figure.
 
 Finally, to close our discussion we show  in Fig.~\ref{Dradius1}  
 the heavy-light meson charge radii obtained with the change of $m_c$ while 
 keeping the other parameters of set (E) and (F) fixed. The heavier 
 the charm quark mass becomes, the charge radius of the
 $D^+$ and $D^+_s$ tends to saturate in a decreasing behavior and the main 
 contribution comes from the light quark, which in our model perceives the 
 minimum branch point value, namely, $m_{\bar d(\bar s)}+\mu_{D^+(D^+_s)}-m_{D^+(D^+_s)}$, 
 which is not affected by the increase of the charm mass and its corresponding localization. 
 Such mechanism is particular to the present model, and mimics 
 the realization of the heavy quark limit of QCD, where the center of 
 the confining force is fixed at the position of the
 heavy quark. In our case, such physics is simulated by keeping the 
 lowest branch point fixed. However, it is necessary to
 distinguish that the present model, although baring some properties that QCD 
 dictates to these heavy-light mesons, it does not posses 
 the absolute confinement of the quarks.

\section{Summary}\label{summary}

In the present work, the  electroweak properties of light and charmed 
$D$ and $D_s$ pseudoscalar mesons 
are investigated within a unified covariant constituent quark model, where
the quark-antiquark-meson vertices are assumed to have  a symmetric form 
by the exchange of quark momenta, 
which were successful in describing the light pseudoscalar meson
properties~\cite{deMelo:2002yq,Yabusaki_2015}. The model
has constituent quark masses up, down, strange and charm, which 
 embody dynamical chiral symmetry breaking and the Higgs contribution 
to the masses. In addition,  each meson has one mass regulator parameter
tuned to reproduce the weak decay constant. 
Consistently with the infrared relevance of the 
dynamical chiral symmetry breaking, we chose a light quark constituent mass 
that reproduces the pion charge radius, even loosing fit at large momentum
of the experimental form factor data.

The reason for such a choice was to study the kaon, $D^+$, $D^+_s$ 
electromagnetic form factors at   low space-like momentum 
transfer squared up to about 1~GeV$^2$, 
where experimental data for the kaon is available~\cite{Amendolia:1986ui,Dally:1980dj}, 
as well as lattice calculations for 
the $D^+$~\cite{Can2012tx,2017EPJA} 
and $D^+_s$~\cite{2017EPJA}. In particular, we gave 
attention 
to the SU(3) and SU(4) flavor symmetry breaking as a consequence of the Higgs 
contribution to the quark masses, where from the LQCD calculations also 
provide the flavor decomposition of these heavy-light meson form factors.
Furthermore, we have  compared our charge radius results with some models from the
literature, which can be useful for contextualizing our effort within what 
has been done. The charge distribution of $\bar s$ is somewhat more 
compact than the corresponding one for the $u$, a feature that 
will be strongly emphasized for the charmed heavy-light pseudoscalars.

Each of the form factors were decomposed in its flavor content,
$F_{u \bar{d} u}(q^2) $ and $ F_{\bar{d} u \bar{d}}(q^2)$
for the pion,  $F_{u \bar{s} u}(q^2)$ and $F_{\bar{s} u \bar{s}}(q^2)$
for the kaon, $F_{c \bar{d} c}(q^2)  $ and 
$F_{\bar{d} c \bar{d}}(q^2)$ for $D^+$,  $F_{c \bar{s} c }(q^2)$ 
and $ F_{\bar{s} c \bar{s}}(q^2)$ for the $D^+_s$. Each of these
flavor form factors has particular properties with respect to the flavor symmetry.
The model has SU(2) flavor symmetry and trivially we 
have $F_{u \bar{d} u}(q^2) = F_{\bar{d} u \bar{d}}(q^2)= F_{\pi^+}(q^2)$,
which we used constituent quark mass that reproduced quite well the pion charge 
radius, favoring a better description of the infrared properties of the model.
The SU(3) flavor symmetry is slightly broken within the kaon
expressed by $F_{\bar{s} u \bar{s}}(q^2)\gtrsim F_{u \bar{s} u}(q^2)$ by 
about 10\% above 3~GeV$^2$, as consequence of the Higgs contribution 
to the strange quark, which makes it about 30\% heavier than the up quark 
in our model. The study of the flavor contribution  to the $D^+$ and $D^+_s$ form 
factors within the model,  showed that the charm quark is localized close
to the center of the meson, 
while the light quark forms a "halo" exploring larger distances where the
confining force of QCD is active, and the infrared physics dominating. 
Such strong asymmetry is a direct consequence of the Higgs providing a
large mass to the charm quark in comparison with light 
ones. The flavor decomposition of the charge radii provided by the model 
is consistent with the lattice QCD calculation from 
Refs.~\cite{,2017EPJA}
where the ensemble C1 was used, and the $D^+$ and $D_s^+$ masses were 
found close to the experimental values. 

We found that the heavier is the charm quark, the charge radius of the
 $D^+$ and $D^+_s$ decrease and saturates, as  dominant contribution to the
 charge radius comes form the light quark. The contribution of the light quark
 to the heavy-light  form factor is essentially dependent on 
 the minimum branch point
 value at  $m_{\bar d(\bar s)}+\mu_{D^+(D^+_s)}-m_{D^+(D^+_s)}$, which is insensitive
 to the increase of the charm quark mass, and its corresponding localization. 
 Such feature is particular to the present model, and is able to incorporate 
 aspects of the heavy quark limit  of QCD, where the force center of the confining
 force stays at the meson center of mass, and therefore the charge distribution 
 becomes independent of the heavy quark mass.
 
 However, the present model does not posses the absolute confinement of 
 the quarks, and they are represented as bound states of constituents quarks. 
 The SU(4) flavor symmetry is largely broken as manifested in the
 contributions to the form factors, where we found in the case of the $D^+$ that
 $F_{c \bar{d} c}(q^2) >F_{\bar{d} c \bar{d}}(q^2)$ and saturating the ratio above
 3~GeV$^2$, attaining  the ratio a value $ \sim 5$. For the  $D^+_s$,  we observe the 
 same behavior as for the $D^+$ having 
 that $F_{c \bar{s} c }(q^2)> F_{\bar{s} c \bar{s}}(q^2)$
 and the ratio saturating at a value around 5 above 3~GeV$^2$.
 This is a strong manifestation of the SU(4) flavor symmetry breaking in 
 the structure of these heavy-light pseudoscalar mesons. We also found that 
 the flavor decomposition of the $D^+$ and $D^+_s$ form factors are in 
 agreement with the lattice QCD calculation from Refs.~\cite{2017EPJA}  
 where the ensemble C1 was used, with results in the space-like region up to
 1.2~GeV$ ^2$.  The structure of the $D^+$ and $D^+_s$ mesons can be studied 
 in much more detail within this model, like for generalized parton distributions, 
 generalized transverse momentum distributions and finally fragmentation functions 
 of the heavy quarks, left for future investigations as well as the application 
 of the model to study $B,\eta_c, J/\psi, \eta_b$ and $\Upsilon$ mesons.

{\it Acknowledgements:}~
This work was supported in part by CAPES, and
by the Conselho Nacional de Desenvolvimento 
Cient\'{i}fico e Tecnol\'{o}gico (CNPq), Grants, No.~308486/2015-3 (TF), 
Process, No.~307131/2020-3 (JPBCM), 
Process, No.~313063/2018-4 (KT), and No.~426150/2018-0 (KT),
and Funda\c{c}\~{a}o de Amparo \`{a} Pesquisa do Estado
de S\~{a}o Paulo (FAPESP), Process, No. 2019/02923-5 (JPBCM), 
No.~2019/00763-0 (KT), and was also part of the projects, Instituto Nacional de Ci\^{e}ncia e
Tecnologia -- Nuclear Physics and Applications (INCT-FNA), Brazil,
Process No.~464898/2014-5, and FAPESP Tem\'{a}tico, Brazil, Process,
the thematic projects, No. 2013/26258-4 and No. 2017/05660-0. 
\\ 

\onecolumngrid

\bibliography{meson.bib} 

\end{document}